# Speckle Statistics of Biological Tissues in Optical Coherence Tomography

**Gary R. Ge,[1] Jannick P. Rolland,[1,3,4] and Kevin J. Parker[2,3,*]**

[1] The Institute of Optics, University of Rochester, Rochester, NY 14627

[2] Department of Electrical and Computer Engineering, University of Rochester, Rochester, NY 14627

[3] Department of Biomedical Engineering, University of Rochester, NY 14627

[4] Center for Visual Science, University of Rochester, NY 14627

* Corresponding author: kevin.parker@rochester.edu

**Abstract:** The speckle statistics of optical coherence tomography images of biological tissue have been studied using several historical probability density functions. A recent hypothesis implies that underlying power-law distributions in the medium structure, such as the fractal branching vasculature, will contribute to power-law probability distributions of speckle statistics. Specifically, these are the Burr type XII distribution for speckle amplitude, the Lomax distribution for intensity, and the generalized logistic distribution for log amplitude. In this study, these three distributions are fitted to histogram data from nine optical coherence tomography scans of various biological tissues and samples. The distributions are also compared with conventional distributions such as the Rayleigh, K, and gamma distributions. The results indicate that these newer distributions based on power laws are, in general, more appropriate models and support the plausibility of their use for characterizing biological tissue. Potentially, the governing power-law parameter of these distributions could be used as a biomarker for tissue disease or pathology.

1.  **Introduction**

Speckle is a granular pattern seen in signals or images that is caused by the interference of coherent waves with random phases and known or random amplitudes.[1] The study of speckle phenomena has a long history dating back to the time of Isaac Newton, with an increasing multitude of recent applications in optics, radar, and ultrasound. Optical coherence tomography (OCT) and ultrasound are two medical imaging modalities with prominent speckle. For some applications such as high-resolution imaging, speckle is considered to be undesirable noise and many studies attempt to eliminate its presence.[2-10] Other studies choose to utilize speckle for physical modeling or characterization of tissue samples.[11-16] Studies of speckle amplitude statistics in acoustics and optics have led to the usage of various probability density functions (PDFs) such as the Rayleigh distribution, the K distribution, the Rice distribution, gamma distributions, and many others.[17-21]

Although OCT is an interferometric technique and ultrasound utilizes time-of-flight measurements, the mathematics describing wave propagation and wave phenomena such as speckle can be applicable to both acoustical and optical imaging modalities. Recently, a new model hypothesized that in normal soft tissue, the dominant scattering elements are cylinders from fractal branching vasculature.[22-25] As a result of this model containing governing power-law relationships, three new and distinct probability distributions were used to characterize ultrasound speckle in biological tissue. These were the Burr type XII distribution, the Lomax distribution, and the generalized logistic distribution.

In this paper, these distributions are extended to OCT scans of various biological tissues. Metrics for assessing appropriate regions of interest (ROIs) and evaluating the statistical validity of the distributions are also presented. Finally, a comparison of the new distributions with distributions found in the literature are presented based on various samples.

## 2. Theory

### 2.1   Modeling of Biological Tissue

Parker et al.[25] derived the first order speckle statistics of biological tissue in ultrasound imaging under the assumptions of weak scattering (using the Born approximation) originating from fractal branching of vasculature represented by cylinders. This derivation leads to power-law functions that dictate the PDFs for the echo amplitude and intensity histograms. The framework for the derivation is also directly applicable to OCT and is summarized here.

First, consider a distribution of scattering structures, from large to small, within the volume. We assume the distribution follows a power-law distribution in size, with fewer larger scatterers. A power law distribution and spatial correlation function are also consistent with generalized fractal models.[26] Thus, in scanning a volume, the probability of encountering a scatterer of characteristic dimension $a$ is given as[25,27]

$$p(a) = \frac{b-1}{a_{\min}} \left(\frac{a}{a_{\min}}\right)^{-b} \quad (1)$$

where $b$ is the power-law coefficient representing the multi-scale nature of the tissue structures, and $a_{\min}$ represents the minimum size or lower limit of dimensions of the scattering structures that are detectable. Previous studies have shown that variations in the index of refraction within tissues obey a power law down to the sub-micron scale.[28] Thus, this model is appropriate for tissues.

Secondly, we assume each scatterer of dimension $a$ produces a detected amplitude $A$ or intensity $I$ according to the theory of backscattered waves. In general, canonical scattering elements such as spheres and cylinders have been characterized by power series solutions.[29-31] The dependence of backscatter on frequency and dimension is complicated, but can be characterized by well-known long-wavelength, short-wavelength, and transition or Mie scattering regimes.[32]

However, we have employed a linear first-order approximation covering the sub-resolvable region where

$$I(a) = I_0(a - a_{min}) \qquad (2)$$

where both $I_0$ and $a_{min}$ are dependent on system parameters such as wavelength and gain, and the lower limit of system detectability, including quantization and noise floor. With this linear monotonic function, the probability of occurrence can be simply mapped into the probability of amplitude or intensity using the probability transformation rule.[23,25]

## 2.2 Probability Distributions for Amplitude, Intensity, and Log of Amplitude

Given Equations 1 and 2, the PDF for the histogram of amplitudes $x$ is determined as

$$p(x; \lambda, b) = \frac{2x(b-1)}{\lambda^2 \left[\left(\frac{x}{\lambda}\right)^2 + 1\right]^b}; \quad x > 0 \qquad (3)$$

which is a special case of the Burr type XII distribution, and $\lambda$ is a normalization parameter. When using this PDF for fitting OCT speckle amplitude, it is convenient to normalize by setting $x = \frac{A}{\sqrt{\langle A^2 \rangle}}$, where the denominator represents the root mean square (RMS) value.[18,21] Otherwise, the normalization constant can be incorporated into the $\lambda$ parameter.

The PDF for the histogram of intensities is given by

$$p(x; \lambda, b) = \frac{(b-1)\lambda^{b-1}}{(x+\lambda)^b}; \quad x > 0 \qquad (4)$$

which is the Lomax distribution or Pareto type II distribution. When using this PDF for fitting OCT speckle intensity $I = |A|^2$, it is convenient to normalize by setting $x = \frac{I}{\langle I \rangle}$, where $\langle I \rangle$ is the mean intensity.

Finally, the PDF for the histogram of log amplitude defined by $y = \ln(A)$ is given by

$$p(y; \lambda, b) = \frac{2(b-1)e^{2y}}{\lambda^2 \left[\frac{e^{2y}}{\lambda^2} + 1\right]^b}; \quad -\infty < y < \infty \tag{5}$$

which is a transformed version of the generalized logistic type I distribution. Typically, in OCT, it is beneficial to display an image of the log of the amplitude or intensity to better visualize the dynamic range. Thus, this distribution is useful in capturing the histograms of most conventional display values. Furthermore, these three PDFs are all well characterized in the statistics and econometrics literature, with known cumulative distribution functions and moments.[33]

### 2.3 The Theoretical Importance of the Exponent Parameter

The above three PDFs all contain a power law or exponent parameter $b$. In power law and related functions, the exponent parameter is a valuable parameter of interest in many applications.[27] In this paper's context, the exponent parameter $b$ may be important in tissue characterization.

According to Carroll-Nellenback, et al.,[34] a simple fractal distribution of vessels within normal tissues would provide a value of approximately $b = 2.7$. However, the number of scatterers per sample volume, and possibly the index of refraction can increase the exponent parameter. Thus, the exponent parameter may provide information about the tissue's scattering properties and structure, and may serve as a biomarker for differentiating disease and pathology from normal.

### 2.4 Historical Probability Distributions for Comparison

In the literature, there are other PDFs used to model OCT speckle amplitudes and intensities based on consideration of random point scatterers or more complex distributions from radar and other fields. The most prevalent of these is the Rayleigh distribution for speckle amplitude and the exponential distribution for speckle intensity, which are given by:[18,20]

$$p(A) = \frac{2A}{\sqrt{\langle A^2 \rangle}} \exp\left(-\frac{A^2}{\langle A^2 \rangle}\right); \quad A > 0 \tag{6}$$

$$p(I) = \frac{1}{\langle I \rangle} \exp\left(-\frac{I}{\langle I \rangle}\right); \quad I > 0 \tag{7}$$

The Rayleigh distribution is suitable for the case of a large number of scatterers in a homogeneous medium.

Another distribution used by Weatherbee *et al.* is the K distribution, which has also been explored in ultrasound and radar.[18,21] The K distribution is modeled for the case of a small number of scatterers, and the PDFs for the amplitudes and intensities are given by

$$p(A; \alpha) = \frac{4}{\Gamma(\alpha)} \sqrt{\frac{\alpha}{\langle A^2 \rangle}} \left(\frac{\alpha A^2}{\langle A^2 \rangle}\right)^{\frac{\alpha}{2}} K_{\alpha-1}\left(2\sqrt{\frac{\alpha A^2}{\langle A^2 \rangle}}\right); \quad A > 0 \tag{8}$$

$$p(I; \alpha) = \frac{2}{\Gamma(\alpha)} \sqrt{\frac{\alpha}{I \langle I \rangle}} \left(\frac{\alpha I}{\langle I \rangle}\right)^{\frac{\alpha}{2}} K_{\alpha-1}\left(2\sqrt{\frac{\alpha I}{\langle I \rangle}}\right); \quad I > 0 \tag{9}$$

where $\Gamma(\cdot)$ is the gamma function and $K(\cdot)$ is a modified Bessel function of the second kind, and $\alpha$ is the shape parameter.

A third distribution for comparison is the gamma distribution, as used by Kirillin *et al.* for modeling speckle amplitude, and is given by[16]

$$p(A; \alpha, \beta) = \frac{1}{\Gamma(\alpha)} \beta^\alpha A^{\alpha-1} e^{-\beta A}; \quad A > 0 \tag{10}$$

where $\alpha$ and $\beta$ are two shape parameters. The PDF for intensity can be derived and is given by

$$p(I; \alpha, \beta) = \frac{1}{2\Gamma(\alpha)} \beta^\alpha I^{\alpha/2 - 1} e^{-\beta \sqrt{I}}; \quad I > 0 \tag{11}$$

## 3. Methods

### 3.1 OCT Scans of Various Biological Tissue

A swept source OCT (SS-OCT) system is used to scan various biological tissue. It is implemented with a swept source laser (HSL-2100-WR, Santec, Aichi, Japan) with a center wavelength of 1310 nm and full-width half-maximum (FWHM) bandwidth of 170 nm. The lateral resolution is approximately 20 $\mu$m and the axial resolution is approximately 8 $\mu$m. The SS-OCT system is controlled with LabVIEW (Version 14, National Instruments, Austin, Texas, USA).

The following tissues were scanned and analyzed with the SS-OCT system: mouse brain and liver, pig brain and cornea, and chicken muscle all *ex vivo* as well as human hand (skin) *in vivo*. In addition, two gelatin phantoms (5% with and without milk for optical scattering) were also scanned and analyzed. The number of A-lines for each scan was either 100, 500, or 1000. Variations in the number of A-lines do not change the speckle statistics, as long as an adequate number of pixels are used (e.g. greater than 1,000 pixels, which is easily covered by a 10 × 100 ROI) over an appropriate field of view (e.g. 5-10 mm). Amplitude, intensity, and log amplitude histograms are generated from specific ROIs in these samples.

*3.2    Fitting Distributions using Maximum Likelihood Estimation*

The distributions specified in Sections 2.2 and 2.4 are fitted to respective amplitude and intensity histograms using maximum likelihood estimation (MLE). The iterative maximization algorithm for MLE and all other analysis aspects were conducted in MATLAB 2020b (MathWorks, Natick, Massachusetts, USA). Curve fitting using an alternative least-squares approach or related methods can result in systemic errors, especially when estimating a power-law or exponent parameter.[35] This is due to a combination of factors such as violations of the underlying assumptions when using these curve fitting methods and variability from histogram binning methods. Therefore, MLE is a more accurate approach to this paper's studies.

*3.3    Metric for Specifying an Appropriate Region of Interest*

Relative uniformity across the ROI is an important, yet difficult to quantify, requirement for assessing appropriate speckle statistics. Attenuation along depth and shadowing effects are two examples of phenomena that would reduce the validity of an ROI for appropriate speckle statistics.

In this subsection, a simulation was performed in order to obtain an appropriate metric for evaluating the selection of an ROI. In Figure 1(a), a simulated OCT B-mode image ($500 \times 500$ pixels) using a Burr type XII distribution with parameters $\lambda = 10$ and $b = 3$ is shown with attenuation and shadowing effects. Figure 1(b) shows how spatial integration profiles vary along the axial and lateral directions, which can be visually correlated to the B-mode image. Specifically, integrations of speckle amplitude are performed along the two directions using intervals of 5 pixels and are normalized so that the two profiles can be displayed on the same scale. Using a Monte Carlo method, 10,000 ROIs of the B-mode image are randomly generated with a minimal size of $15 \times 15$ pixels to ensure adequate statistics. The resulting speckle statistics were fitted to a Burr type XII distribution using MLE. The ROIs were quantified by the maximum percent change (i.e. ratio of maximum difference over minimum value) in their spatial integration profiles, Δ. Figure 1(c) shows how the estimated parameter $\hat{b}$ varies as a function of the ROI's Δ. Using this simulation, we established an *ad hoc* requirement that any appropriate ROI's Δ should not exceed a 50% change in their spatial integration profiles laterally and axially.

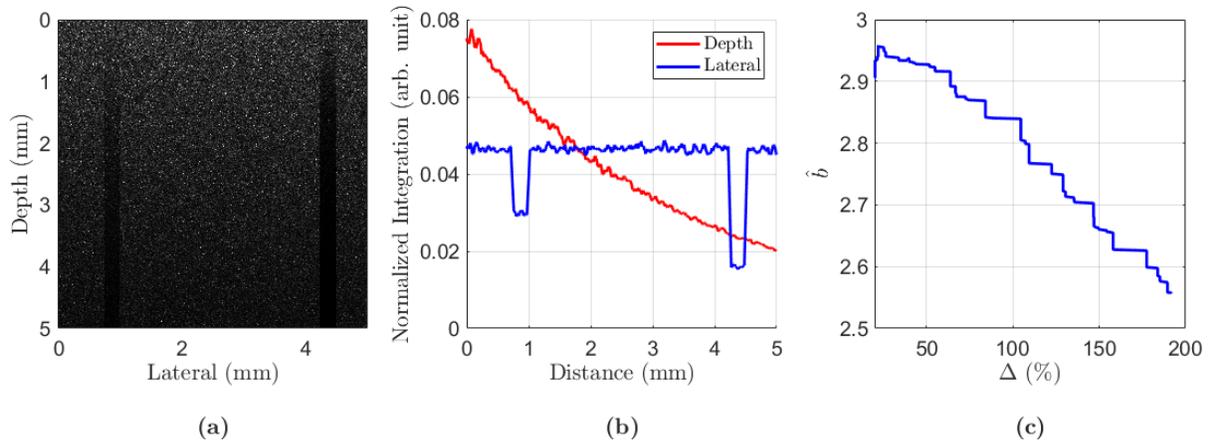

**Figure 1:** Determining an ROI metric. **(a)** Simulated B-mode image using a Burr type XII distribution ($\lambda = 10, b = 3$) with attenuation and shadowing. **(b)** Depth and lateral profiles based on spatial integration laterally and along depth, respectively. **(c)** Multiple ROIs and MLE fitting were used to calculate $\hat{b}$ as a function of maximum percent change in profiles, Δ.

*3.4    Comparing Multiple Distributions*

There are numerous methods for direct comparison of models. These methods include the Kolmogorov-Smirnov (KS) test, the Anderson-Darling (AD) test, the likelihood ratio test (LRT), and the Akaike information criterion (AIC).[35-37] The KS and AD test statistics provide *p*-values and indicate which models are best rejected while simultaneously indicating which model is the best fit. The AD test is similar to the KS test but gives more weight to distribution tails. However, it can be too conservative with estimates of power law functions and requires customized tests for the PDFs described in Sections 2.2 and 2.4. Instead, we follow the rapid KS procedure as outlined in Clauset *et al.*[35] By performing the KS test on the sample data and simulated sample distributions from the MLE fit, a *p*-value can be estimated by using a Monte Carlo method. In this case, a very large *p*-value ($> 0.9$) indicates a good fit, while a small value indicates that the model is not an appropriate one. When comparing models, the one with the largest KS *p*-value can be considered the best, although the best model may still not be a good fit.

The second test used is the LRT. The LRT generates a ratio of the log likelihoods of two models for comparison, which we denote by $R$.[35,36] In this case, the LRT is used to compare the distributions in Section 2.4 (Rayleigh/Exponential, K, and Gamma distributions) with the Burr/Lomax distributions. The sign of $R$ indicates which distribution is the better fit: if $R < 0$, then the Burr/Lomax distribution is the preferred distribution. A *p*-value is also estimated with $R$ to acknowledge that a true $R = 0$ may fluctuate either positively or negatively. If the *p*-value is small, then the sign of $R$ is a good indicator of which distribution is the better fit. Otherwise for a non-small *p*-value ($> 0.1$), the LRT is inconclusive.

The final metric used is the AIC, which is defined as

$$AIC = 2k - 2\ln\left(\hat{L}\right) \quad (12)$$

where $k$ is the number of estimated parameters and $\hat{L}$ is the model's likelihood. The AIC provides a relative measure of a model when compared to other models, where a smaller value indicates a better model.[37] These three methods (KS test, LRT, and AIC) are used to determine which distribution fits best for speckle amplitude (Burr type XII, Rayleigh, K, or Gamma) and intensity (Lomax, Exponential, K, or Gamma).

## 4. Results

### 4.1 *Evaluating the Burr type XII, Lomax, and Generalized Logistic Distributions*

Figure 2 demonstrates the speckle analysis performed for a mouse brain sample. Figure 2(a) shows the B-mode image and the ROI used for the speckle statistics, and Figure 2(b) verifies that this ROI is appropriate using the metric defined in Section 3.3. The MLE fits for the amplitude values using the Burr type XII distribution are shown in Figure 2(c)-(d), on both a linear and logarithmic scale. Figure 2(e)-(f) reports the MLE fits for the intensity values using the Lomax distribution, and Figure 2(g)-(h) shows the MLE fits for the log of amplitude values using the generalized logistic distribution. All MLE fits to the prescribed PDFs result in a consistent exponent parameter estimate $\hat{b} = 6.06$ (95% Confidence Interval: [5.83, 6.28]) for mouse brain. Figure 3 demonstrates the same type of analysis performed for mouse liver, which results in a consistent estimate of $\hat{b} = 5.95$ (95% Confidence Interval: [5.70, 6.21]). Visual inspection indicates that the Burr type XII, the Lomax, and the generalized logistic distributions are good fits to the data, but statistical analyses of these distributions are further investigated in the next section.

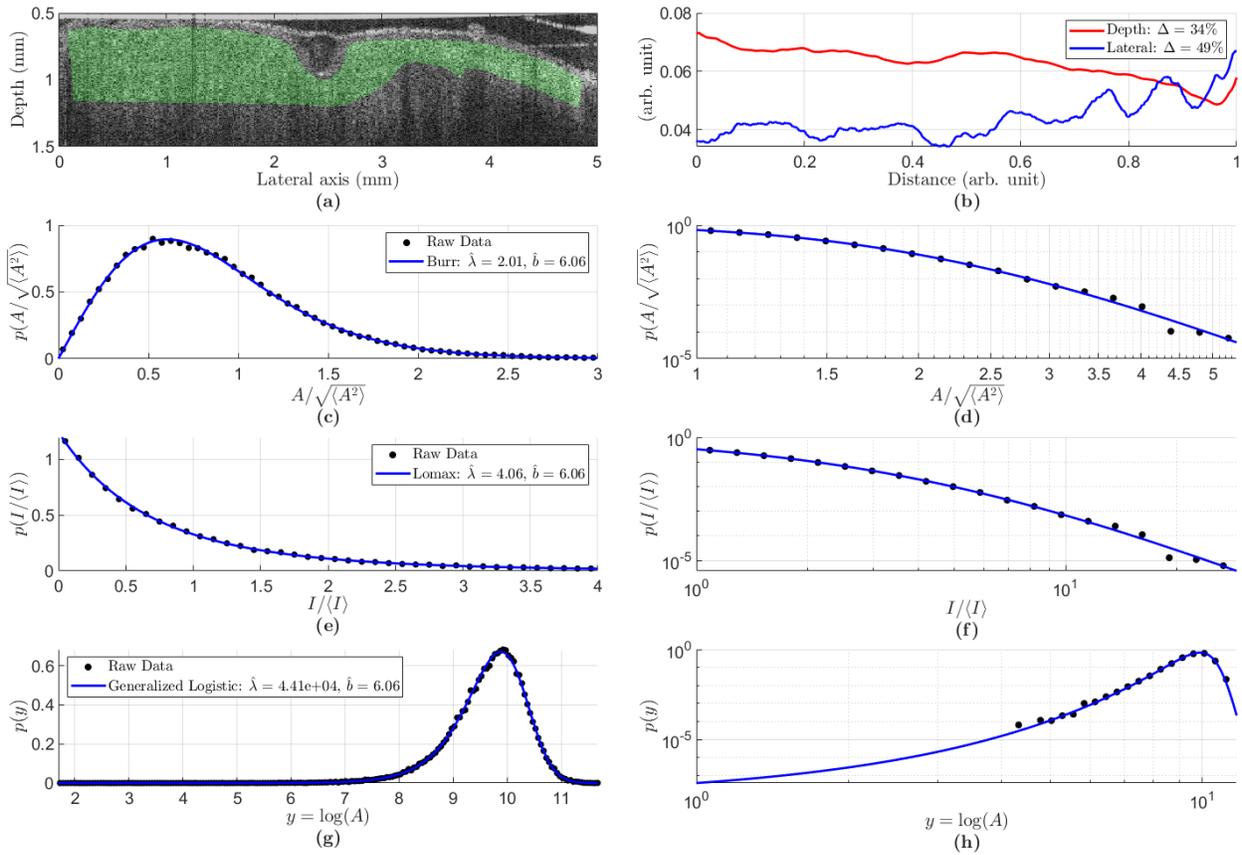

**Figure 2:** Mouse brain with cranial window. **(a)** Single unfiltered and unaveraged B-mode frame with the ROI shaded green. **(b)** The ROI's normalized spatial integration profiles demonstrating that the maximum percent change Δ does not exceed 50%. **(c)** MLE fit to the Burr type XII distribution in a linear scale with linear binning of amplitude histogram data. **(d)** Same as (c) but in a loglog scale with logarithmic binning for visualization of tail behavior. **(e)** MLE fit to the Lomax distribution in a linear scale with linear binning of intensity histogram data. **(f)** Same as (d) but in a log-log scale with logarithmic binning to visualize tail behavior. **(g)** MLE fit to the generalized logistic distribution in a linear scale with linear binning of log amplitude histogram data. **(h)** Same as (g) but in log-log scale with logarithmic binning. All MLE fits result in an exponent parameter of $\hat{b} = 6.06$ (95% Confidence Interval: [5.83, 6.28]) for mouse brain.

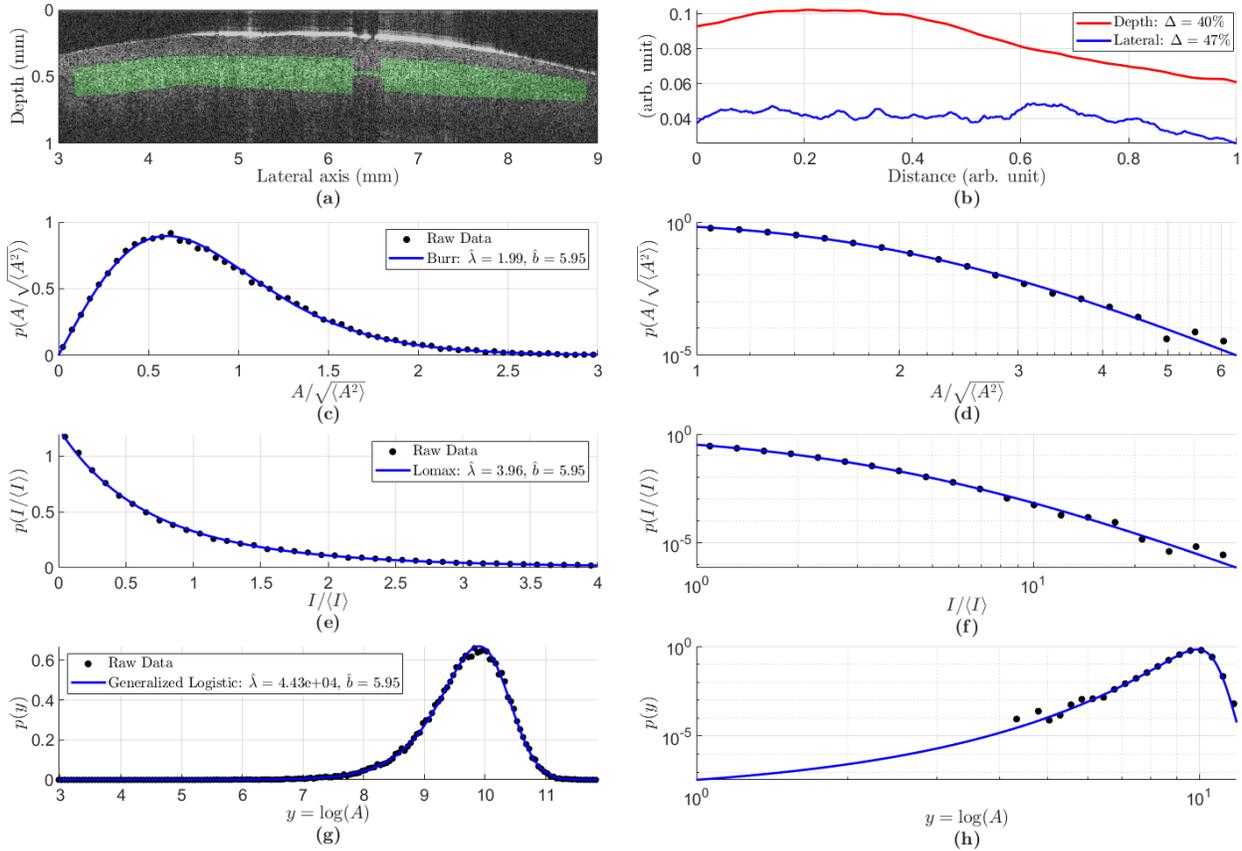

**Figure 3:** Excised mouse liver in phosphate-buffered saline. **(a)** Single unfiltered and unaveraged B-mode frame with the ROI shaded green. **(b)** The ROI's normalized spatial integration profiles demonstrating that the maximum percent change Δ does not exceed 50%. **(c)** MLE fit to the Burr type XII distribution in a linear scale with linear binning of amplitude histogram data. **(d)** Same as (c) but in a loglog scale with logarithmic binning for visualization of tail behavior. **(e)** MLE fit to the Lomax distribution in a linear scale with linear binning of intensity histogram data. **(f)** Same as (d) but in a log-log scale with logarithmic binning to visualize tail behavior. **(g)** MLE fit to the generalized logistic distribution in a linear scale with linear binning of log amplitude histogram data. **(h)** Same as (g) but in log-log scale with logarithmic binning. All MLE fits result in an exponent parameter of $\hat{b} = 5.95$ (95% Confidence Interval: [5.70, 6.21]) for mouse liver.

*4.2    Comparing Multiple Distributions*

Figure 4 demonstrates multiple distribution fits for a pig cornea sample. Figure 4(a) shows the B-mode frame for the pig cornea with an appropriate ROI as quantified in Figure 4(b). Figure 4(c)-(d) depicts the MLE fits for amplitude histogram data using the Burr type XII, the Rayleigh, the K, and the gamma distributions as specified in Section 2.4 in both a linear and logarithmic scale. The intensity histogram data and the Lomax, exponential, K, and gamma distribution MLE fits are shown in Figure 4(e)-(f). All respective MLE fits for amplitude and intensity result in consistent parameter estimates (e.g. $\hat{\alpha}$ and $\hat{\beta}$ are the same for the gamma distributions for both amplitude and intensity). Figure 5 shows the sample analysis done for a human hand's backside. Once again, the MLE parameter estimates are consistent between amplitude and intensity fits for each pair of distributions (Burr/Lomax, Rayleigh/exponential, K, or gamma). The complete set of data for all tissue or phantom samples taken can be found in Supplemental Figures S1-S9. In some cases, on the logarithmic scale, there are slight deviations between the data points and the estimated Burr and Lomax distributions. Since there are a finite number of pixels in a single ROI, there is a minimum probability estimate that can be captured by histogram data. Additionally, the bins in the tail carry a small number of samples, and so statistical fluctuations account for a larger proportion of these samples. Hence, results in the tail become noisy, and deviations from the estimated distributions are expected.

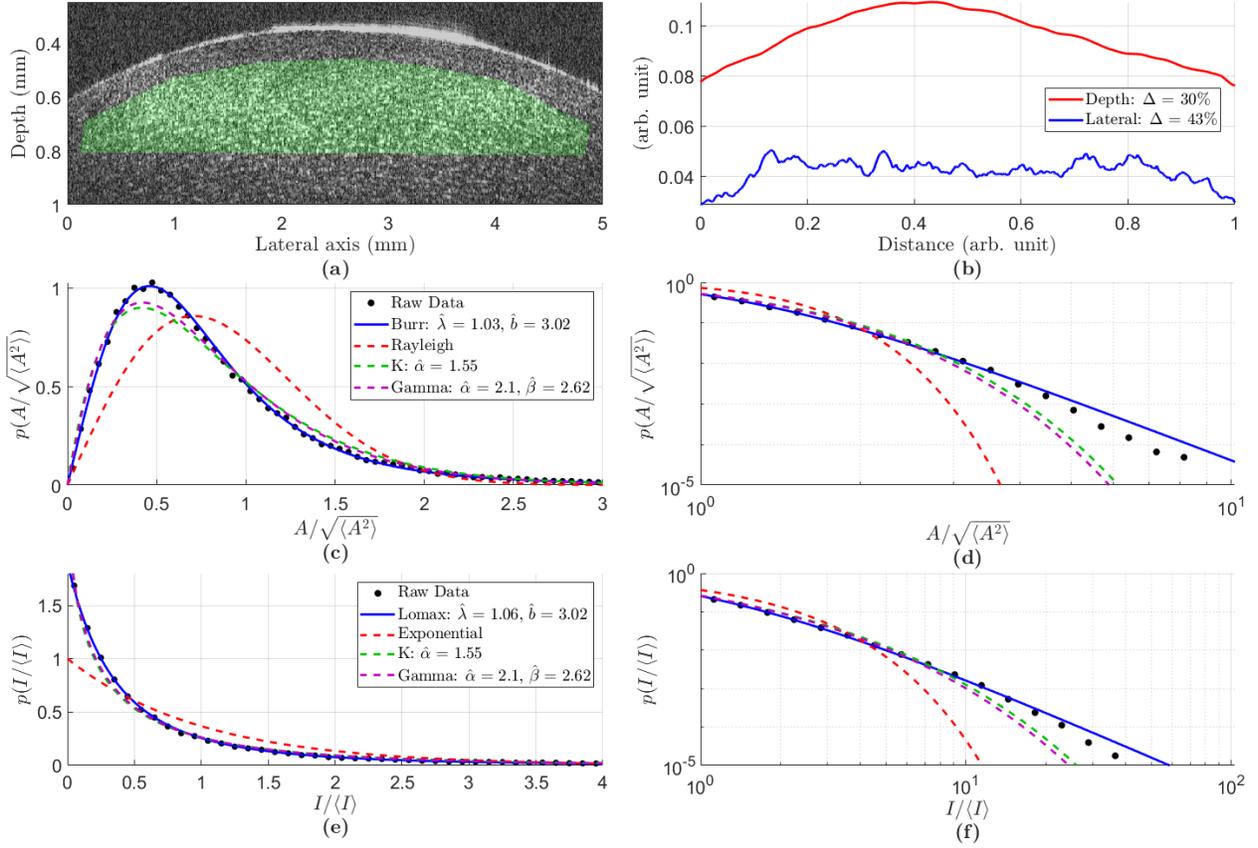

**Figure 4:** Pig cornea. **(a)** Single unfiltered and unaveraged B-mode image with the ROI shaded green. **(b)** The ROI's normalized spatial integration profiles that do not exceed a Δ of 50%. **(c)** MLE fits to normalized amplitude with the Burr, Rayleigh, K, and Gamma distributions on a linear scale with linear binning. **(d)** Same as part (c) except on a log-log scale with logarithmic binning to visualize tail behavior. **(e)** MLE fits to normalized intensity with the Lomax, Exponential, K, and Gamma distributions on a linear scale with linear binning. **(f)** Same as part (e) except on a log-log scale with logarithmic binning to visualize tail behavior.

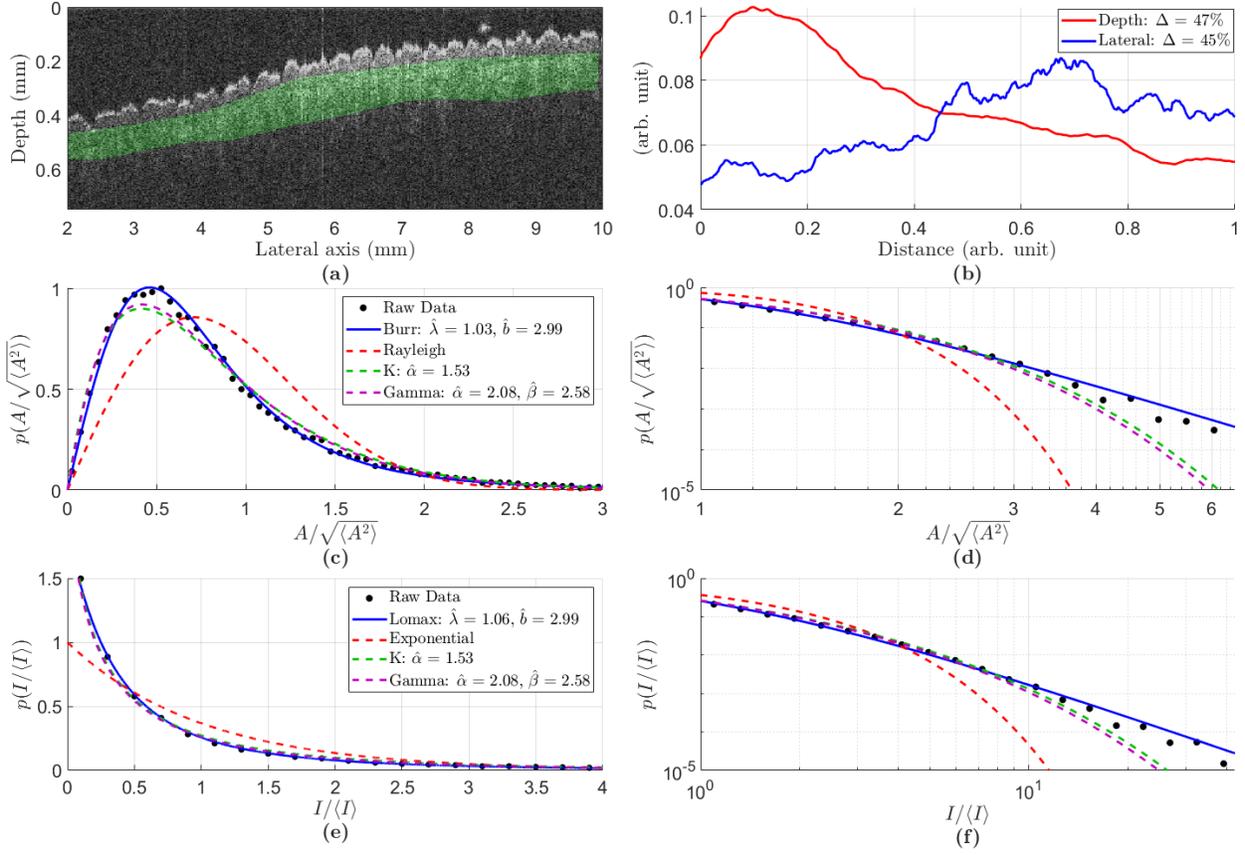

**Figure 5:** Human hand (backside). **(a)** Single unfiltered and unaveraged B-mode image with the ROI shaded green. **(b)** The ROI's normalized spatial integration profiles which do not exceed a Δ of 50%. **(c)** MLE fits to normalized amplitude with the Burr, Rayleigh, K, and Gamma distributions on a linear scale with linear binning. **(d)** Same as part (c) except on a log-log scale with logarithmic binning to visualize tail behavior. **(e)** MLE fits to normalized intensity with the Lomax, Exponential, K, and Gamma distributions on a linear scale with linear binning. **(f)** Same as part (e) except on a log-log scale with logarithmic binning to visualize tail behavior.

The next step is to perform the statistical tests and metrics for comparing multiple distributions described in Section 3.4. The results of all data samples are detailed in Supplemental Tables S1-S9, but a summary table of the comparison outcomes is provided below in Supplemental Table S10. In all samples except for one, the best model for amplitude was the Burr type XII distribution and the best model for intensity was the Lomax distribution. The one exception is for pig brain, in which the K distribution was the best fit for both amplitude and intensity data.

*4.3    Estimating the Exponent Parameter of Multiple Types of Biological Tissues*

The Burr type XII, Lomax, and generalized logistic distributions all have two parameters: $\lambda$ and $b$. $\lambda$ can be seen as a system-dependent (and gain-dependent) normalization parameter, but the exponent parameter $b$ is of interest since it could be used to characterize tissue. For all samples, $\hat{b}$ and its 95% confidence interval were obtained via MLE in Supplemental Figures S1-S9. A summary of the exponent values for various samples types are shown in Table 1, in order of increasing $\hat{b}$. Note that for all samples except the pig brain, the Burr/Lomax distribution was determined to be the best model in the previous section.

**Table 1**: Estimated exponent parameter values for all samples.

| Sample | $\hat{b}$ | 95% Confidence Interval |
|---|---|---|
| Human Hand (Backside) | 2.99 | [2.94, 3.05] |
| Pig Cornea | 3.02 | [2.98, 3.06] |
| Human Hand (Palm) | 3.12 | [3.09, 3.16] |
| 5% Gelatin Phantom | 3.54 | [3.44, 3.64] |
| Chicken Muscle | 4.59 | [4.38, 4.80] |
| 5% Gelatin Phantom (+ Milk) | 5.83 | [5.70, 5.97] |
| Pig Brain | 5.91 | [5.18, 6.64] |
| Mouse Liver | 5.95 | [5.70, 6.21] |
| Mouse Brain | 6.06 | [5.83, 6.28] |

## 5. Discussion and Conclusion

In Section 4.1, MLE fits to the Burr type XII, Lomax, and generalized logistic distributions show that these three PDFs fit reasonably well to the amplitude, intensity, and log amplitude histogram data for two sample tissues (mouse brain and liver). The estimated exponential parameter $\hat{b}$ is also universally consistent due to the framework of MLE. Thus, any one of the three

PDFs can be used to estimate the exponential parameter $b$ with a reasonable level of confidence. In Section 4.2, we further demonstrate the merits of using these PDFs (Burr/Lomax), as they are statistically compared with conventional PDFs described in the literature (Rayleigh/exponential, K, and gamma). Out of nine total samples, eight of them demonstrated that Burr/Lomax are indeed the best fits for the amplitude and intensity histogram data. In the one exception (pig brain), the differences between the Burr/Lomax fits and the K fits were statistically significant but relatively small. Another trend that was noticed was the fact that if the Burr/Lomax estimated a relatively high $\hat{b}$ (~ 6), then the K and gamma fits were relatively closer to the Burr/Lomax fits, and they could also be a reasonable fit to the data (this can be verified by checking that the KS *p*-values are large). On the other hand, if the Burr/Lomax $\hat{b}$ was small (~3), then the K and gamma distributions were poorer fits to the data (small KS *p*-values). Overall, the Burr/Lomax distributions indicate that they are reliable and generalizable to characterizing the speckle statistics of various biological tissue in OCT.

In Section 4.3, the exponent parameter $\hat{b}$ is tabulated and compared across multiple biological tissues. The range for $\hat{b}$ is from ~3 to ~6. There may be some reasonable explanations for the general trend of these values. The samples with lower values of $\hat{b}$ ~3 are all tissues that are relatively transparent to light (eye, skin, gelatin), whereas the samples with higher values of $\hat{b}$ ~6 are considered optically denser tissues (liver, brain). The differences between the phantom studies also supports this concept. The gelatin phantom with milk (an added optical scatterer) has a higher $\hat{b}$ value. Furthermore, since the Burr/Lomax distributions are much better fits to the speckle statistics than Rayleigh/K/gamma distributions for tissues with low values of $\hat{b}$, then the Burr/Lomax may be even more important in characterizing these types of tissues. OCT currently has strong clinical applications in ophthalmology and dermatology, and so a framework involving $\hat{b}$ could provide a useful biomarker for pathology or disease.

In summary, this study has presented three new PDFs for fitting speckle statistics in various biological tissues: the Burr type XII distribution for amplitude, the Lomax distribution for intensity, and the generalized logistic distribution for log of amplitude. Furthermore, an *ad hoc* metric for verifying an appropriate ROI, the MLE fitting technique, and methodology for statistical comparison of multiple distributions are also presented. Future work is needed to closely link the speckle parameter *b* with spatial measures of tissue microstructure in 3D.[28,34,38] These independent measures on a larger set of tissues will be needed to verify that the results of this study are reproducible, and to determine the importance of the exponent parameter *b* in multiple biological and clinical applications.

**Acknowledgments**

Gary Ge is supported by the National Institute on Aging of the National Institutes of Health under award number F30AG069293. The work is also supported by NIH grants R21EB025290 and R21AG070331. The content is solely the responsibility of the authors and does not necessarily represent the official views of the National Institutes of Health. The authors would also like to thank Fernando Zvietcovich for providing the OCT scans of pig brain and chicken muscle.

**Author Contributions**

G.R.G., J.P.R., and K.J.P. conceived and designed the project; G.R.G. performed the acquisition and analysis of the data. G.R.G and K.J.P. interpreted the data. G.R.G., J.P.R., and K.J.P. wrote and revised the manuscript.

**Competing Interests**

The authors have no competing interests to declare.

**Data Availability**

All relevant data and codes are available from the authors.

Material requests and correspondence should be addressed to K.J.P.

## References

1 Goodman, J. W. *Speckle Phenomena in Optics: Theory and Applications*. 2 edn, (SPIE The International Society for Optics and Photonics, 2020).

2 Bashkansky, M. & Reintjes, J. Statistics and reduction of speckle in optical coherence tomography. *Opt. Lett.* **25**, 545-547, doi:10.1364/OL.25.000545 (2000).

3 Abd-Elmoniem, K. Z., Youssef, A. M. & Kadah, Y. M. Real-time speckle reduction and coherence enhancement in ultrasound imaging via nonlinear anisotropic diffusion. *IEEE Transactions on Biomedical Engineering* **49**, 997-1014, doi:10.1109/TBME.2002.1028423 (2002).

4 Michael, P., Erich, G., Rainer, A. L., Adolf Friedrich, F. & Christoph, K. H. Speckle reduction in optical coherence tomography by frequency compounding. *Journal of Biomedical Optics* **8**, 565-569, doi:10.1117/1.1578087 (2003).

5 Adler, D. C., Ko, T. H. & Fujimoto, J. G. Speckle reduction in optical coherence tomography images by use of a spatially adaptive wavelet filter. *Opt. Lett.* **29**, 2878-2880, doi:10.1364/OL.29.002878 (2004).

6 Kim, J. *et al.* Optical coherence tomography speckle reduction by a partially spatially coherent source. *Journal of Biomedical Optics* **10**, 064034 (2005).

7 Ozcan, A., Bilenca, A., Desjardins, A. E., Bouma, B. E. & Tearney, G. J. Speckle reduction in optical coherence tomography images using digital filtering. *J. Opt. Soc. Am. A* **24**, 1901-1910, doi:10.1364/JOSAA.24.001901 (2007).

8 Guo, Y., Cheng, H. D., Tian, J. & Zhang, Y. A Novel Approach to Speckle Reduction in Ultrasound Imaging. *Ultrasound in Medicine & Biology* **35**, 628-640, doi:https://doi.org/10.1016/j.ultrasmedbio.2008.09.007 (2009).

9 Park, J., Kang, J. B., Chang, J. H. & Yoo, Y. Speckle reduction techniques in medical ultrasound imaging. *Biomedical Engineering Letters* **4**, 32-40, doi:10.1007/s13534-014-0122-6 (2014).

10 Liba, O. *et al.* Speckle-modulating optical coherence tomography in living mice and humans. *Nature Communications* **8**, 15845, doi:10.1038/ncomms15845 (2017).

11 Tuthill, T. A., Sperry, R. H. & Parker, K. J. Deviations from Rayleigh statistics in ultrasonic speckle. *Ultrasonic Imaging* **10**, 81-89, doi:https://doi.org/10.1016/0161-7346(88)90051-X (1988).

12 Thijssen, J. M. Ultrasonic speckle formation, analysis and processing applied to tissue characterization. *Pattern Recognition Letters* **24**, 659-675, doi:https://doi.org/10.1016/S0167-8655(02)00173-3 (2003).

13 Karamata, B., Hassler, K., Laubscher, M. & Lasser, T. Speckle statistics in optical coherence tomography. *J. Opt. Soc. Am. A* **22**, 593-596, doi:10.1364/JOSAA.22.000593 (2005).
20

14	Di, L., Rao, N., Kuo, C., Bhatt, S. & Dogra, V. in *2007 29th Annual International Conference of the IEEE Engineering in Medicine and Biology Society.*  6523-6526.

15	Lindenmaier, A. A. *et al.* Texture analysis of optical coherence tomography speckle for characterizing biological tissues in vivo. *Opt. Lett.* **38**, 1280-1282, doi:10.1364/OL.38.001280 (2013).

16	Kirillin, M. Y., Farhat, G., Sergeeva, E. A., Kolios, M. C. & Vitkin, A. Speckle statistics in OCT images: Monte Carlo simulations and experimental studies. *Opt. Lett.* **39**, 3472-3475, doi:10.1364/OL.39.003472 (2014).

17	Al-Kadi, O. S., Chung, D. Y. F., Coussios, C. C. & Noble, J. A. Heterogeneous Tissue Characterization Using Ultrasound: A Comparison of Fractal Analysis Backscatter Models on Liver Tumors. *Ultrasound in Medicine & Biology* **42**, 1612-1626, doi:https://doi.org/10.1016/j.ultrasmedbio.2016.02.007 (2016).

18	Weatherbee, A., Sugita, M., Bizheva, K., Popov, I. & Vitkin, A. Probability density function formalism for optical coherence tomography signal analysis: a controlled phantom study. *Opt. Lett.* **41**, 2727-2730, doi:10.1364/OL.41.002727 (2016).

19	Kalkman, J. Fourier-Domain Optical Coherence Tomography Signal Analysis and Numerical Modeling. *International Journal of Optics* **2017**, 9586067, doi:10.1155/2017/9586067 (2017).

20	Almasian, M., van Leeuwen, T. G. & Faber, D. J. OCT Amplitude and Speckle Statistics of Discrete Random Media. *Scientific Reports* **7**, 14873, doi:10.1038/s41598-017-14115-3 (2017).

21	Stanton, T. K., Lee, W.-J. & Baik, K. Echo statistics associated with discrete scatterers: A tutorial on physics-based methods. *The Journal of the Acoustical Society of America* **144**, 3124-3171, doi:10.1121/1.5052255 (2018).

22	Parker, K. J. Shapes and distributions of soft tissue scatterers. *Physics in Medicine & Biology* **64**, doi:https://doi.org/10.1088/1361-6560/ab2485 (2019).

23	Parker, K. J. The first order statistics of backscatter from the fractal branching vasculature. *The Journal of the Acoustical Society of America* **146**, 3318-3326, doi:10.1121/1.5132934 (2019).

24	Parker, K. J. & Poul, S. S. Speckle from branching vasculature: dependence on number density. *Journal of Medical Imaging* **7**, 1-12, doi:10.1117/1.JMI.7.2.027001 (2020).

25	Parker, K. J. & Poul, S. S. Burr, Lomax, Pareto, and Logistic Distributions from Ultrasound Speckle. *Ultrasonic Imaging* **42**, 203-212, doi:10.1177/0161734620930621 (2020).

26	Vicsek, T. *Fractal Growth Phenomena*.  (WORLD SCIENTIFIC, 1992).

27	Newman, M. E. J. Power laws, Pareto distributions and Zipf's law. *Contemporary Physics* **46**, 323-351, doi:10.1080/00107510500052444 (2005).

28  Schmitt, J. M. & Kumar, G. Turbulent nature of refractive-index variations in biological tissue. *Opt. Lett.* **21**, 1310-1312, doi:10.1364/OL.21.001310 (1996).

29  Rayleigh, L. X. On the electromagnetic theory of light. *The London, Edinburgh, and Dublin Philosophical Magazine and Journal of Science* **12**, 81-101, doi:10.1080/14786448108627074 (1881).

30  Mie, G. Beiträge zur Optik trüber Medien, speziell kolloidaler Metallösungen. *Annalen der physik* **330**, 377-445 (1908).

31  Faran Jr, J. J. Sound scattering by solid cylinders and spheres. *The Journal of the Acoustical Society of America* **23**, 405-418 (1951).

32  Born, M. & Wolf, E. *Principles of Optics: Electromagnetic Theory of Propagation, Interference and Diffraction of Light*. 7 edn, (Cambridge University Press, 1999).

33  McLaughlin, M. P. Compendium of Common Probability Distributions. (2016).

34  Carroll-Nellenback, J. J., White, R. J., Wood, R. W. & Parker, K. J. Liver Backscatter and the Hepatic Vasculature's Autocorrelation Function. *Acoustics* **2**, 3-12 (2020).

35  Clauset, A., Shalizi, C. R. & Newman, M. E. J. Power-Law Distributions in Empirical Data. *SIAM Review* **51**, 661-703, doi:10.1137/070710111 (2009).

36  Vuong, Q. H. Likelihood Ratio Tests for Model Selection and Non-Nested Hypotheses. *Econometrica* **57**, 307-333, doi:10.2307/1912557 (1989).

37  Akaike, H. A new look at the statistical model identification. *IEEE Transactions on Automatic Control* **19**, 716-723, doi:10.1109/TAC.1974.1100705 (1974).

38  Risser, L. *et al.* From Homogeneous to Fractal Normal and Tumorous Microvascular Networks in the Brain. *Journal of Cerebral Blood Flow & Metabolism* **27**, 293-303, doi:10.1038/sj.jcbfm.9600332 (2006).

# Supplemental Figures and Tables

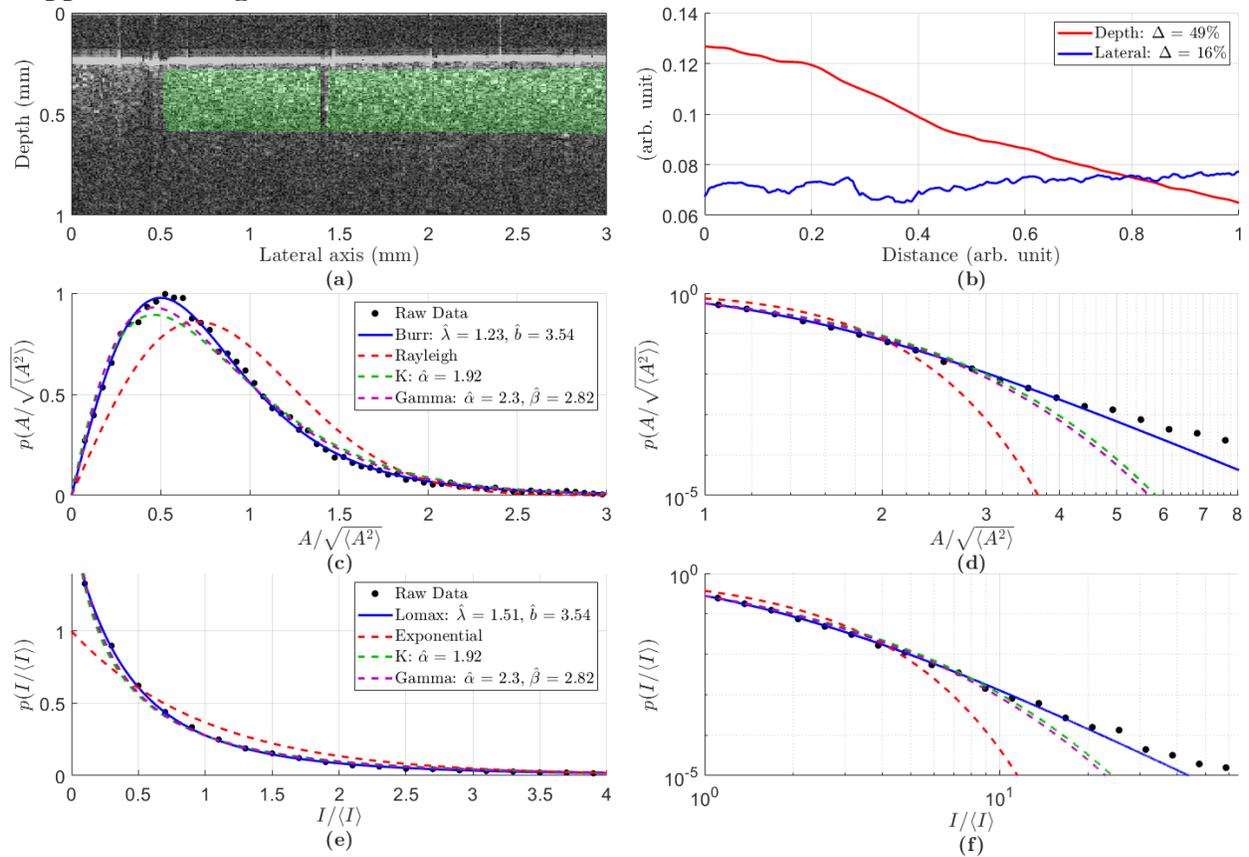

**Figure S1**: 5% gelatin phantom without milk scanned with 500 A-lines. **(a)** Single unfiltered and unaveraged B-mode image with ROI shaded green. **(b)** ROI's normalized spatial integration profiles which do not exceed a Δ of 50%. **(c)** MLE fits to normalized amplitude with the Burr, Rayleigh, K, and Gamma distributions on a linear scale with linear binning. **(d)** Same as part (c) except on a log-log scale with logarithmic binning to visualize tail behavior. **(e)** MLE fits to normalized intensity with the Lomax, Exponential, K, and Gamma distributions on a linear scale with linear binning. **(f)** Same as part (e) except on a log-log scale with logarithmic binning to visualize tail behavior.

**Table S1**: Parameters and statistics for the 5% gelatin phantom without milk.

| Amplitude Distribution | Parameters | Parameters Estimate | KS p-value | LRT $R$ | LRT p-value | AIC | Interpretation |
|---|---|---|---|---|---|---|---|
| Burr type XII | $\hat{\lambda}$ | 1.23 | 0.964 | | | 38823 | - KS, LRT, and AIC indicate Burr is the best model. |
| | $\hat{b}$ | 3.54 | | | | | |
| Rayleigh | | | 0 | -3806 | <0.000001 (1E-6) | 46432 | |
| K | $\hat{\alpha}$ | 1.92 | 0.449 | -396 | <0.000001 (1E-6) | 39614 | |
| Gamma | $\hat{\alpha}$ | 2.30 | 0.762 | -337 | <0.000001 (1E-6) | 39499 | |
| | $\hat{\beta}$ | 2.82 | | | | | |
| **Intensity Distribution** | **Parameters** | **Parameters Estimate** | **KS p-value** | **LRT $R$** | **LRT p-value** | **AIC** | **Interpretation** |
| Lomax | $\hat{\lambda}$ | 1.51 | 0.909 | | | 19255 | - KS, LRT, and AIC indicate Lomax is the best model. |
| | $\hat{b}$ | 3.54 | | | | | |
| Exponential | | | 0 | -22768 | <0.000001 (1E-6) | 64787 | |
| K | $\hat{\alpha}$ | 1.92 | 0.428 | -3143 | <0.000001 (1E-6) | 25540 | |
| Gamma | $\hat{\alpha}$ | 2.30 | 0.720 | -3511 | <0.000001 (1E-6) | 26278 | |
| | $\hat{\beta}$ | 2.82 | | | | | |

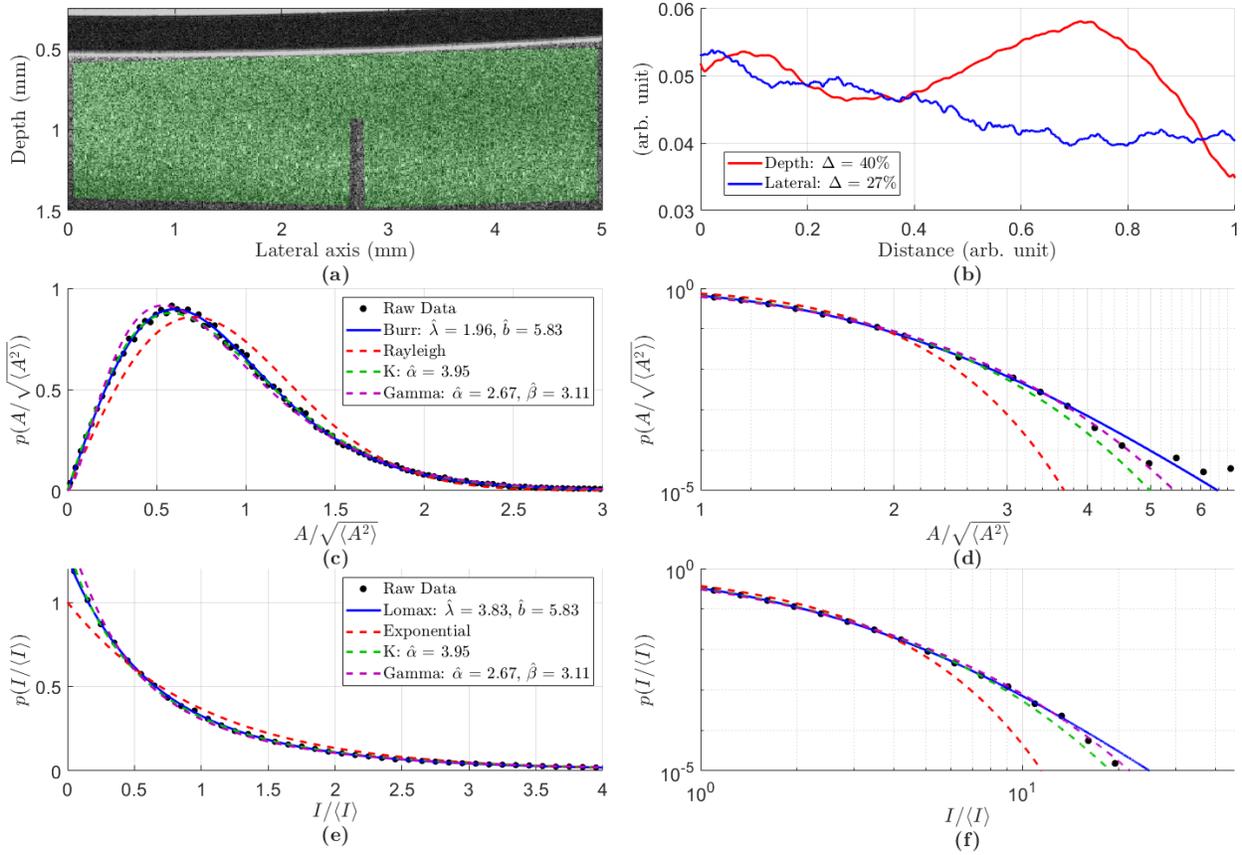

**Figure S2**: 5% gelatin phantom with milk scanned with 500 A-lines. **(a)** Single unfiltered and unaveraged B-mode image with ROI shaded green. **(b)** ROI's normalized spatial integration profiles which do not exceed a Δ of 50%. **(c)** MLE fits to normalized amplitude with the Burr, Rayleigh, K, and Gamma distributions on a linear scale with linear binning. **(d)** Same as part (c) except on a log-log scale with logarithmic binning to visualize tail behavior. **(e)** MLE fits to normalized intensity with the Lomax, Exponential, K, and Gamma distributions on a linear scale with linear binning. **(f)** Same as part (e) except on a log-log scale with logarithmic binning to visualize tail behavior.

**Table S2**: Parameters and statistics for the 5% gelatin phantom with milk.

| Amplitude Distribution | Parameters | Parameters Estimate | KS p-value | LRT $R$ | LRT p-value | AIC | Interpretation |
|---|---|---|---|---|---|---|---|
| Burr type XII | $\hat{\lambda}$ | 1.96 | 0.958 | | | 222575 | - KS, LRT, and AIC indicate Burr is the best model. |
| | $\hat{b}$ | 5.83 | | | | | |
| Rayleigh | | | 0.033 | -4287 | <0.000001 (1E-6) | 231145 | |
| K | $\hat{\alpha}$ | 3.95 | 0.924 | -138 | <0.000001 (1E-6) | 222850 | |
| Gamma | $\hat{\alpha}$ | 2.67 | 0.841 | -476 | <0.000001 (1E-6) | 223528 | |
| | $\hat{\beta}$ | 3.11 | | | | | |
| **Intensity Distribution** | **Parameters** | **Parameters Estimate** | **KS p-value** | **LRT $R$** | **LRT p-value** | **AIC** | **Interpretation** |
| Lomax | $\hat{\lambda}$ | 3.83 | 0.869 | | | 20250 | - KS, LRT, and AIC indicate Lomax is the best model. |
| | $\hat{b}$ | 5.83 | | | | | |
| Exponential | | | 0.002 | -12798 | <0.000001 (1E-6) | 45842 | |
| K | $\hat{\alpha}$ | 3.95 | 0.738 | -2164 | <0.000001 (1E-6) | 24577 | |
| Gamma | $\hat{\alpha}$ | 2.67 | 0.095 | -1052 | <0.000001 (1E-6) | 22355 | |
| | $\hat{\beta}$ | 3.11 | | | | | |

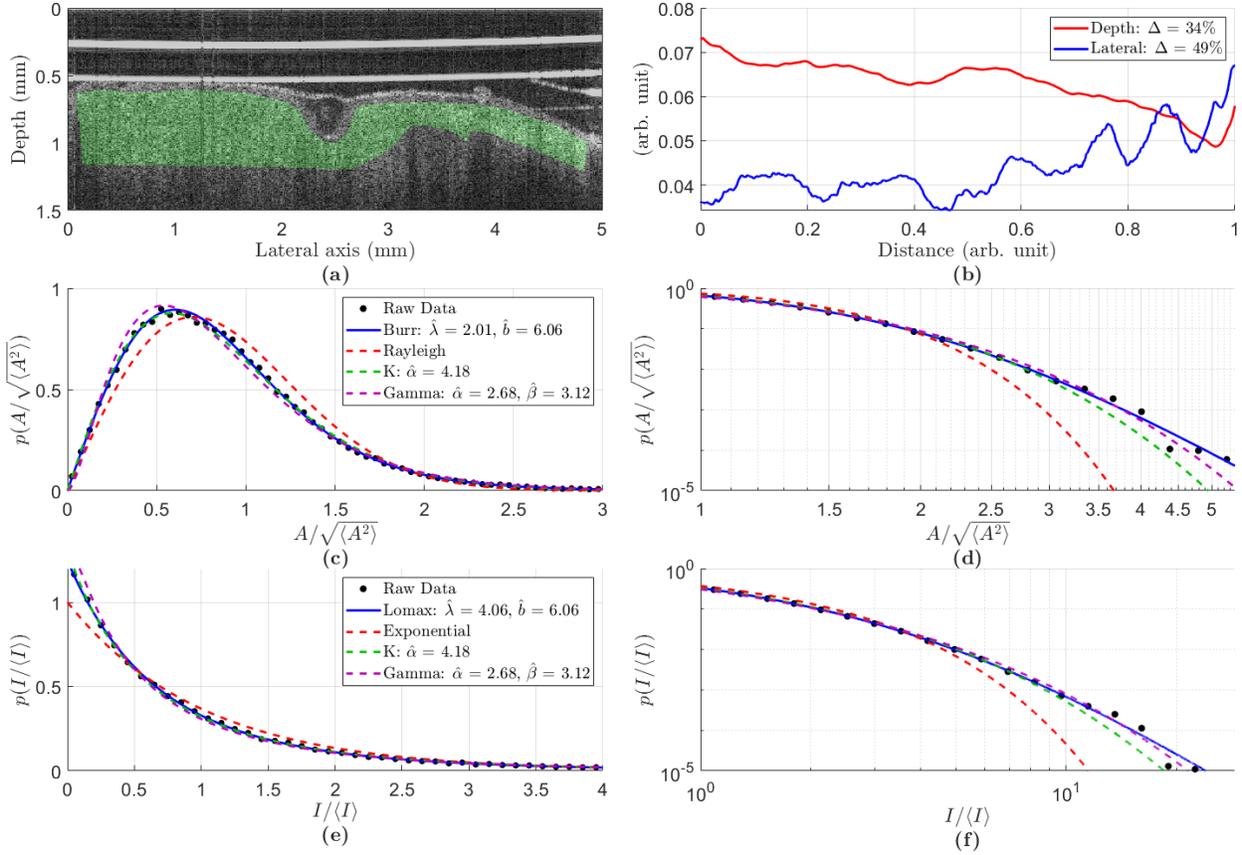

**Figure S3:** Mouse brain with cranial window scanned with 500 A-lines. **(a)** Single unfiltered and unaveraged B-mode image with ROI shaded green. **(b)** ROI's normalized spatial integration profiles which do not exceed a Δ of 50%. **(c)** MLE fits to normalized amplitude with the Burr, Rayleigh, K, and Gamma distributions on a linear scale with linear binning. **(d)** Same as part (c) except on a log-log scale with logarithmic binning to visualize tail behavior. **(e)** MLE fits to normalized intensity with the Lomax, Exponential, K, and Gamma distributions on a linear scale with linear binning. **(f)** Same as part (e) except on a log-log scale with logarithmic binning to visualize tail behavior.

**Table S3**: Parameters and statistics for the mouse brain with cranial window.

| Amplitude Distribution | Parameters | Parameters Estimate | KS p-value | LRT $R$ | LRT p-value | AIC | Interpretation |
|---|---|---|---|---|---|---|---|
| Burr type XII | $\hat{\lambda}$ | 2.01 | 0.949 | | | 91077 | - KS, LRT, and AIC indicate Burr is the best model. |
| | $\hat{b}$ | 6.06 | | | | | |
| Rayleigh | | | 0.075 | -1622 | <0.000001 (1E-6) | 94317 | |
| K | $\hat{\alpha}$ | 4.18 | 0.929 | -67 | <0.000001 (1E-6) | 91210 | |
| Gamma | $\hat{\alpha}$ | 2.68 | 0.816 | -268 | <0.000001 (1E-6) | 91614 | |
| | $\hat{\beta}$ | 3.12 | | | | | |
| **Intensity Distribution** | **Parameters** | **Parameters Estimate** | **KS p-value** | **LRT $R$** | **LRT p-value** | **AIC** | **Interpretation** |
| Lomax | $\hat{\lambda}$ | 4.06 | 0.851 | | | 16400 | - KS, LRT, and AIC indicate Lomax is the best model. |
| | $\hat{b}$ | 6.06 | | | | | |
| Exponential | | | 0.036 | -6280 | <0.000001 (1E-6) | 28956 | |
| K | $\hat{\alpha}$ | 4.18 | 0.828 | -995 | <0.000001 (1E-6) | 18389 | |
| Gamma | $\hat{\alpha}$ | 2.68 | 0.662 | -253 | <0.000001 (1E-6) | 16906 | |
| | $\hat{\beta}$ | 3.12 | | | | | |

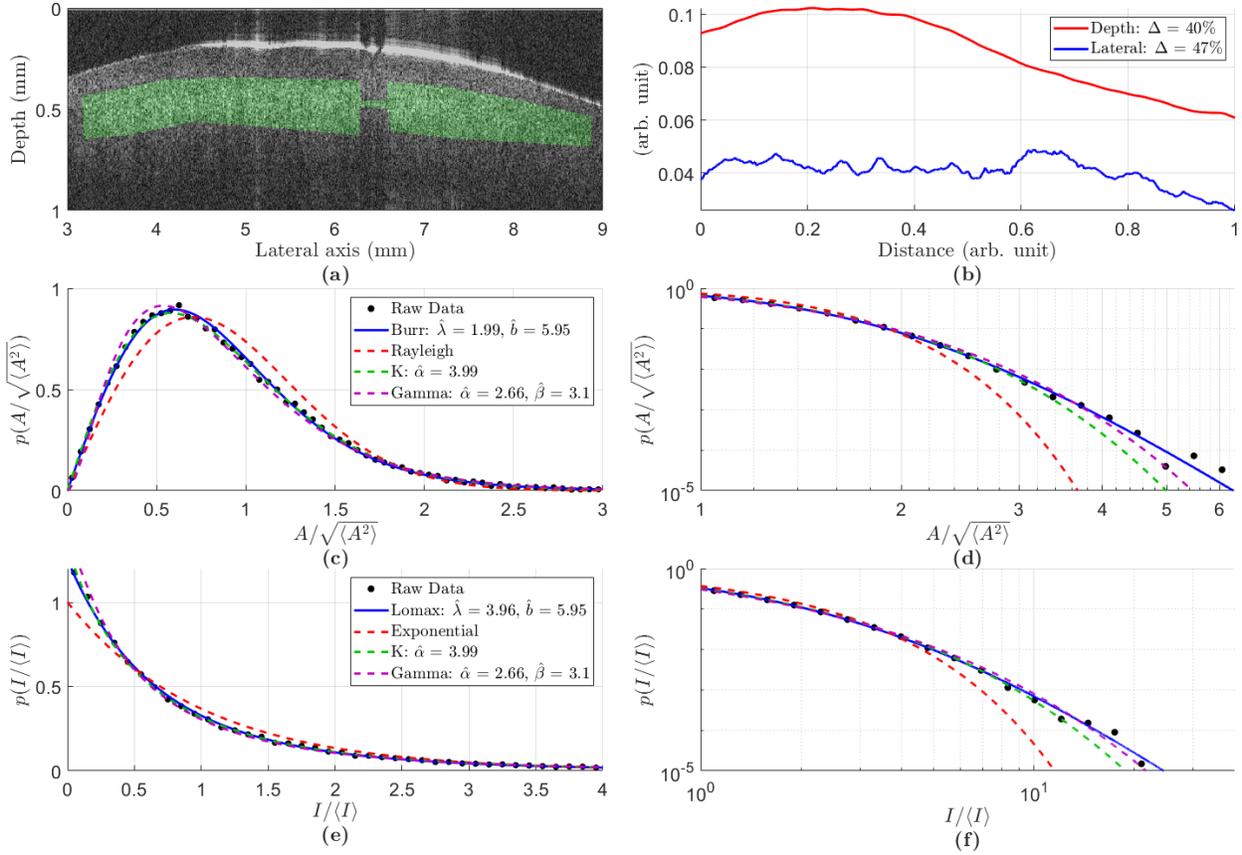

**Figure S4**: Excised mouse liver in phosphate-buffered saline scanned with 1000 A-lines. **(a)** Single unfiltered and unaveraged B-mode image with ROI shaded green. **(b)** ROI's normalized spatial integration profiles which do not exceed a Δ of 50%. **(c)** MLE fits to normalized amplitude with the Burr, Rayleigh, K, and Gamma distributions on a linear scale with linear binning. **(d)** Same as part (c) except on a log-log scale with logarithmic binning to visualize tail behavior. **(e)** MLE fits to normalized intensity with the Lomax, Exponential, K, and Gamma distributions on a linear scale with linear binning. **(f)** Same as part (e) except on a log-log scale with logarithmic binning to visualize tail behavior.

**Table S4**: Parameters and statistics for the mouse liver.

| Amplitude Distribution | Parameters | Parameters Estimate | KS p-value | LRT $R$ | LRT p-value | AIC | Interpretation |
|---|---|---|---|---|---|---|---|
| Burr type XII | $\hat{\lambda}$ | 1.99 | 0.957 | | | 65863 | - KS, LRT, and AIC indicate Burr is the best model. |
| | $\hat{b}$ | 5.95 | | | | | |
| Rayleigh | | | 0.043 | -1181 | <0.000001 (1E-6) | 68220 | |
| K | $\hat{\alpha}$ | 3.99 | 0.935 | -19 | 0.091 | 65898 | |
| Gamma | $\hat{\alpha}$ | 2.66 | 0.879 | -134 | <0.000001 (1E-6) | 66131 | |
| | $\hat{\beta}$ | 3.10 | | | | | |
| **Intensity Distribution** | **Parameters** | **Parameters Estimate** | **KS p-value** | **LRT $R$** | **LRT p-value** | **AIC** | **Interpretation** |
| Lomax | $\hat{\lambda}$ | 3.96 | 0.841 | | | 19147 | - KS, LRT, and AIC indicate Lomax is the best model. |
| | $\hat{b}$ | 5.95 | | | | | |
| Exponential | | | 0.004 | -10424 | <0.000001 (1E-6) | 39990 | |
| K | $\hat{\alpha}$ | 3.99 | 0.771 | -1703 | <0.000001 (1E-6) | 22551 | |
| Gamma | $\hat{\alpha}$ | 2.66 | 0.223 | -4703 | <0.000001 (1E-6) | 24554 | |
| | $\hat{\beta}$ | 3.10 | | | | | |

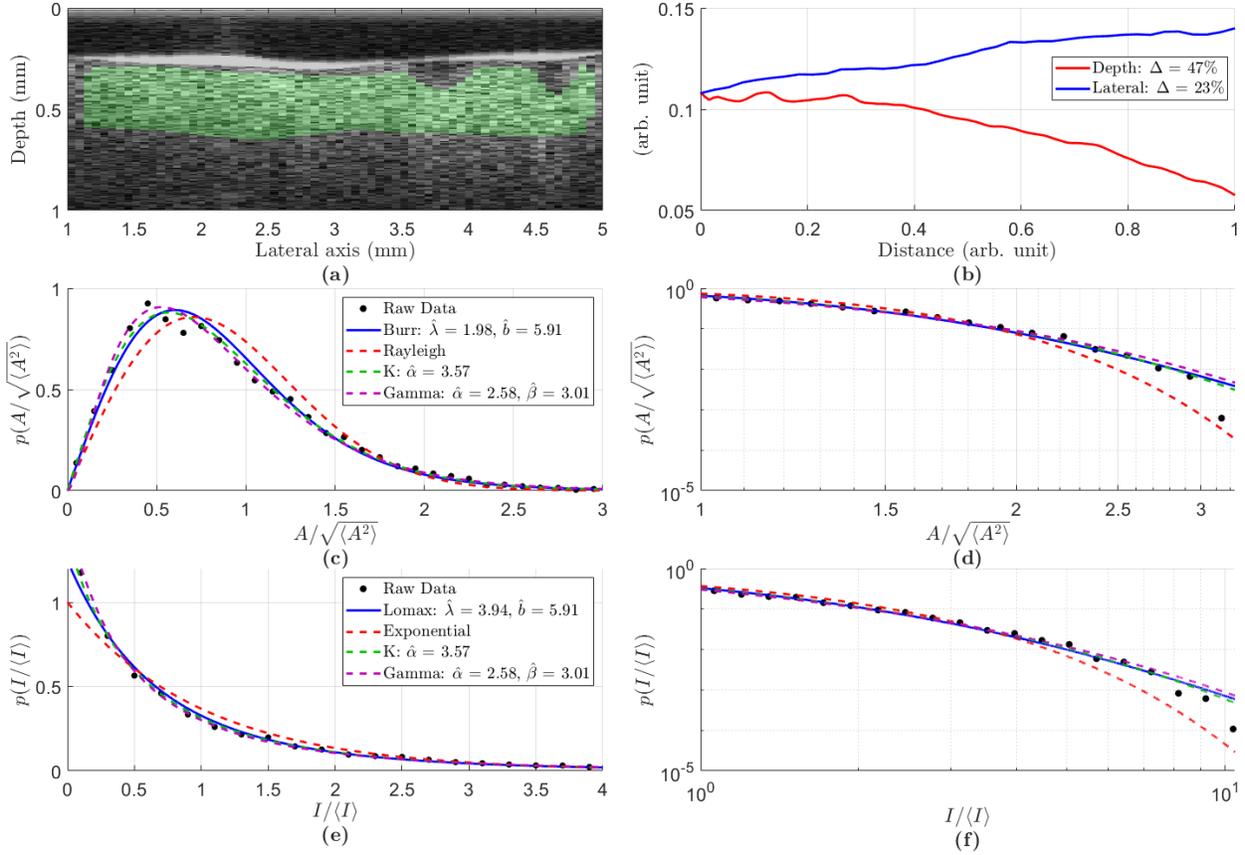

**Figure S5**: Excised pig brain (cortex) scanned with 100 A-lines. **(a)** Single unfiltered and unaveraged B-mode image with ROI shaded green. **(b)** ROI's normalized spatial integration profiles which do not exceed a Δ of 50%. **(c)** MLE fits to normalized amplitude with the Burr, Rayleigh, K, and Gamma distributions on a linear scale with linear binning. **(d)** Same as part (c) except on a log-log scale with logarithmic binning to visualize tail behavior. **(e)** MLE fits to normalized intensity with the Lomax, Exponential, K, and Gamma distributions on a linear scale with linear binning. **(f)** Same as part (e) except on a log-log scale with logarithmic binning to visualize tail behavior.

**Table S5**: Parameters and statistics for the pig brain.

| Amplitude Distribution | Parameters | Parameters Estimate | KS p-value | LRT $R$ | LRT p-value | AIC | Interpretation |
|---|---|---|---|---|---|---|---|
| Burr type XII | $\hat{\lambda}$ | 1.98 | 0.892 | | | 9696 | - KS, LRT, and AIC indicate K is the best model. |
| | $\hat{b}$ | 5.91 | | | | | |
| Rayleigh | | | 0.013 | -125 | <0.000001 (1E-6) | 9941 | |
| K | $\hat{\alpha}$ | 3.57 | 0.943 | 30 | <0.000001 (1E-6) | 9635 | |
| Gamma | $\hat{\alpha}$ | 2.58 | 0.913 | 10 | 0.15 | 9676 | |
| | $\hat{\beta}$ | 3.01 | | | | | |
| **Intensity Distribution** | **Parameters** | **Parameters Estimate** | **KS p-value** | **LRT $R$** | **LRT p-value** | **AIC** | **Interpretation** |
| Lomax | $\hat{\lambda}$ | 3.94 | 0.886 | | | 8824 | - KS, LRT, and AIC indicate K is the best model. |
| | $\hat{b}$ | 5.91 | | | | | |
| Exponential | | | 0.005 | -812 | <0.000001 (1E-6) | 10444 | |
| K | $\hat{\alpha}$ | 3.57 | 0.925 | 55 | 0.024 | 8575 | |
| Gamma | $\hat{\alpha}$ | 2.58 | 0.921 | 12 | <0.000001 (1E-6) | 8834 | |
| | $\hat{\beta}$ | 3.01 | | | | | |

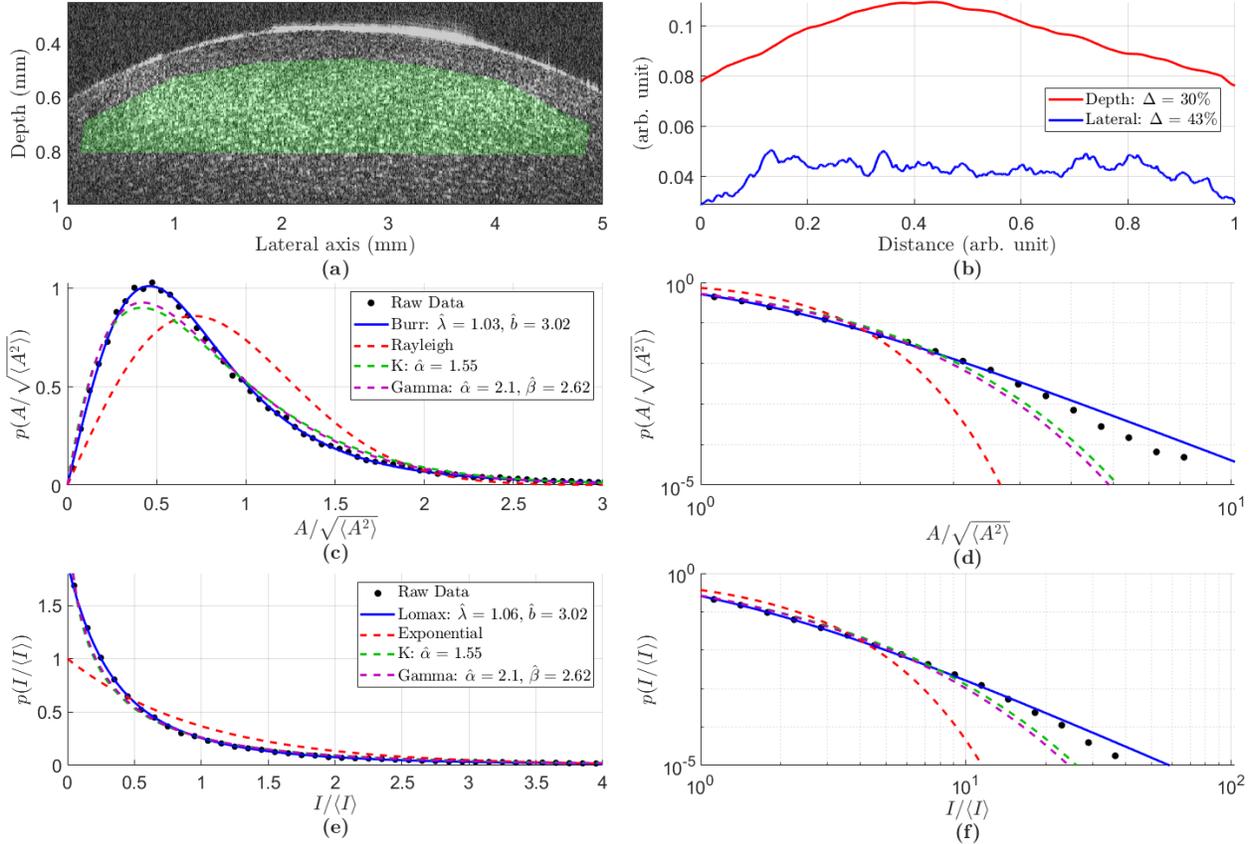

**Figure S6**: Pig cornea scanned with 1000 A-lines. **(a)** Single unfiltered and unaveraged B-mode image with ROI shaded green. **(b)** ROI's normalized spatial integration profiles which do not exceed a Δ of 50%. **(c)** MLE fits to normalized amplitude with the Burr, Rayleigh, K, and Gamma distributions on a linear scale with linear binning. **(d)** Same as part (c) except on a log-log scale with logarithmic binning to visualize tail behavior. **(e)** MLE fits to normalized intensity with the Lomax, Exponential, K, and Gamma distributions on a linear scale with linear binning. **(f)** Same as part (e) except on a log-log scale with logarithmic binning to visualize tail behavior.

**Table S6**: Parameters and statistics for the pig cornea.

| Amplitude Distribution | Parameters | Parameters Estimate | KS p-value | LRT $R$ | LRT p-value | AIC | Interpretation |
|---|---|---|---|---|---|---|---|
| Burr type XII | $\hat{\lambda}$ | 1.03 | 0.958 | | | 134916 | - KS, LRT, and AIC indicate Burr is the best model. |
| | $\hat{b}$ | 3.02 | | | | | |
| Rayleigh | | | 0 | -15291 | <0.000001 (1E-6) | 165495 | |
| K | $\hat{\alpha}$ | 1.55 | 0.391 | -963 | <0.000001 (1E-6) | 136840 | |
| Gamma | $\hat{\alpha}$ | 2.10 | 0.626 | -862 | <0.000001 (1E-6) | 136641 | |
| | $\hat{\beta}$ | 2.62 | | | | | |
| Intensity Distribution | Parameters | Parameters Estimate | KS p-value | LRT $R$ | LRT p-value | AIC | Interpretation |
| Lomax | $\hat{\lambda}$ | 1.06 | 0.944 | | | 20650 | - KS, LRT, and AIC indicate Lomax is the best model. |
| | $\hat{b}$ | 3.02 | | | | | |
| Exponential | | | 0 | -41343 | <0.000001 (1E-6) | 103331 | |
| K | $\hat{\alpha}$ | 1.55 | 0.432 | -5272 | <0.000001 (1E-6) | 31193 | |
| Gamma | $\hat{\alpha}$ | 2.10 | 0.697 | -5984 | <0.000001 (1E-6) | 32619 | |
| | $\hat{\beta}$ | 2.62 | | | | | |

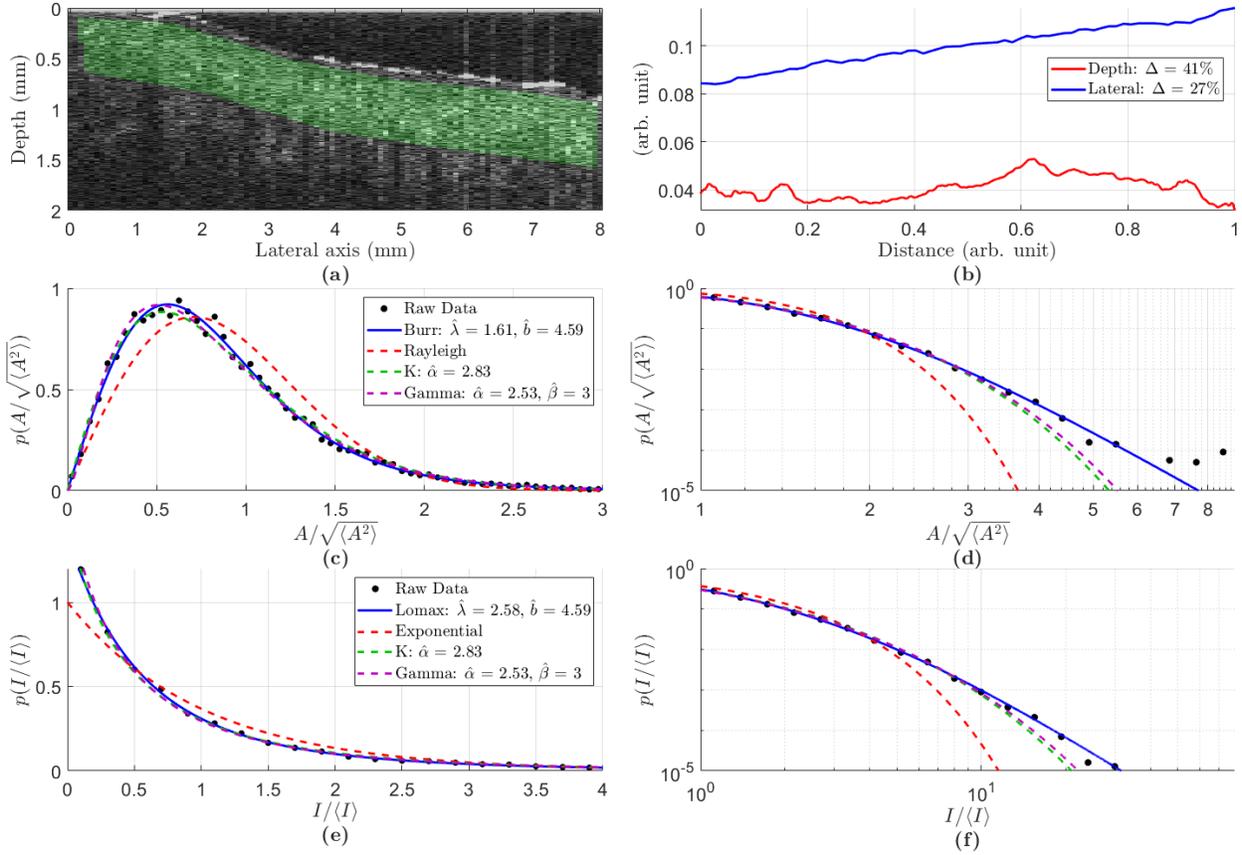

**Figure S7:** Excised chicken muscle scanned with 100 A-lines. **(a)** Single unfiltered and unaveraged B-mode image with ROI shaded green. **(b)** ROI's normalized spatial integration profiles which do not exceed a Δ of 50%. **(c)** MLE fits to normalized amplitude with the Burr, Rayleigh, K, and Gamma distributions on a linear scale with linear binning. **(d)** Same as part (c) except on a log-log scale with logarithmic binning to visualize tail behavior. **(e)** MLE fits to normalized intensity with the Lomax, Exponential, K, and Gamma distributions on a linear scale with linear binning. **(f)** Same as part (e) except on a log-log scale with logarithmic binning to visualize tail behavior.

**Table S7**: Parameters and statistics for the chicken muscle.

| Amplitude Distribution | Parameters | Parameters Estimate | KS p-value | LRT $R$ | LRT p-value | AIC | Interpretation |
|---|---|---|---|---|---|---|---|
| Burr type XII | $\hat{\lambda}$ | 1.61 | 0.952 | | | 30196 | - KS, LRT, and AIC indicate Burr is the best model. |
| | $\hat{b}$ | 4.59 | | | | | |
| Rayleigh | | | 0.001 | -1215 | <0.000001 (1E-6) | 32621 | |
| K | $\hat{\alpha}$ | 2.83 | 0.884 | -73 | 0.00023 | 30340 | |
| Gamma | $\hat{\alpha}$ | 2.53 | 0.899 | -59 | 0.00077 | 30314 | |
| | $\hat{\beta}$ | 3.00 | | | | | |
| **Intensity Distribution** | **Parameters** | **Parameters Estimate** | **KS p-value** | **LRT $R$** | **LRT p-value** | **AIC** | **Interpretation** |
| Lomax | $\hat{\lambda}$ | 2.58 | 0.907 | | | 22988 | - KS, LRT, and AIC indicate Lomax is the best model. |
| | $\hat{b}$ | 4.59 | | | | | |
| Exponential | | | 0 | -28483 | <0.000001 (1E-6) | 79951 | |
| K | $\hat{\alpha}$ | 2.83 | 0.845 | -4438 | <0.000001 (1E-6) | 31863 | |
| Gamma | $\hat{\alpha}$ | 2.53 | 0.876 | -3718 | <0.000001 (1E-6) | 30425 | |
| | $\hat{\beta}$ | 3.00 | | | | | |

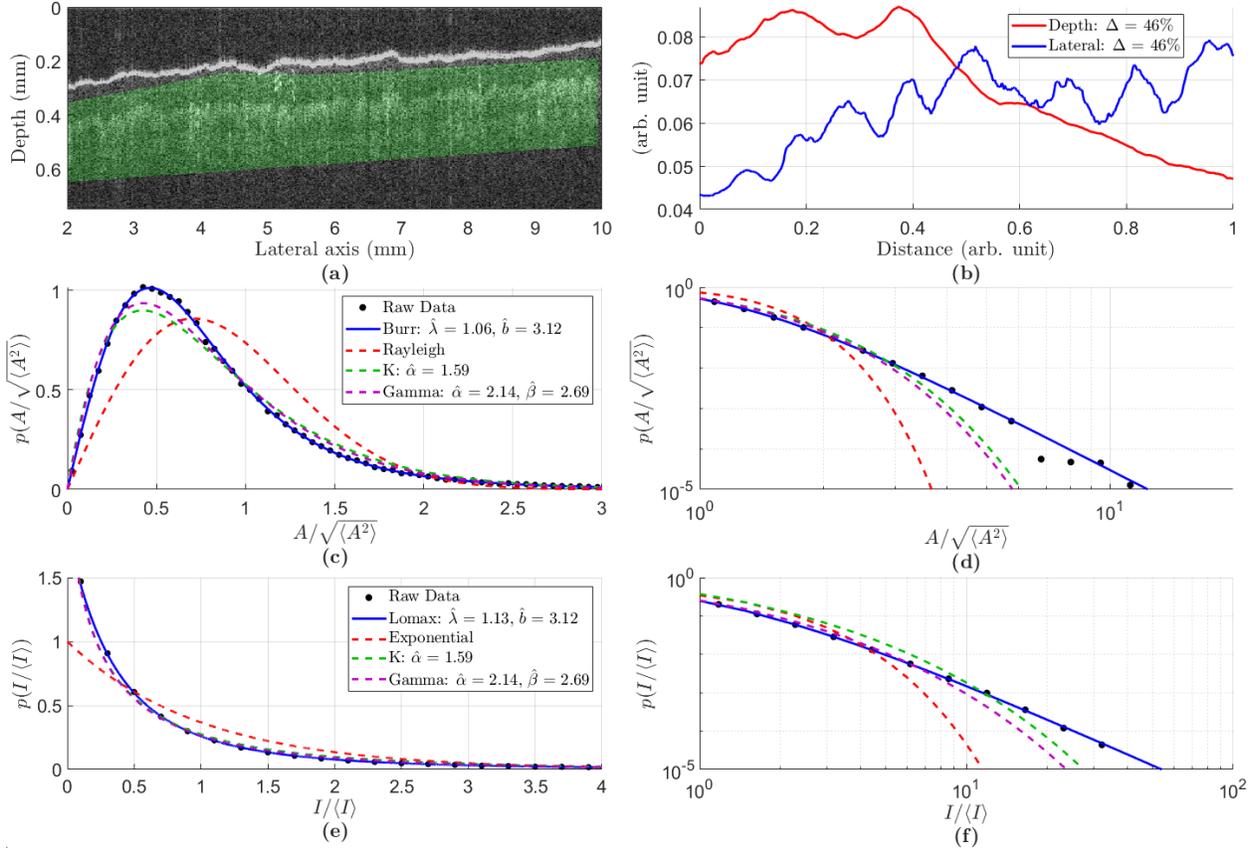

**Figure S8**: Human hand (palm) scanned with 1000 A-lines. **(a)** Single unfiltered and unaveraged B-mode image with ROI shaded green. **(b)** ROI's normalized spatial integration profiles which do not exceed a Δ of 50%. **(c)** MLE fits to normalized amplitude with the Burr, Rayleigh, K, and Gamma distributions on a linear scale with linear binning. **(d)** Same as part (c) except on a log-log scale with logarithmic binning to visualize tail behavior. **(e)** MLE fits to normalized intensity with the Lomax, Exponential, K, and Gamma distributions on a linear scale with linear binning. **(f)** Same as part (e) except on a log-log scale with logarithmic binning to visualize tail behavior.

**Table S8**: Parameters and statistics for the human hand's palm.

| Amplitude Distribution | Parameters | Parameter Estimates | KS p-value | LRT R | LRT p-value | AIC | Interpretation |
|---|---|---|---|---|---|---|---|
| Burr type XII | $\hat{\lambda}$ | 1.06 | 0.894 | | | 157194 | - KS, LRT, and AIC indicate Burr is the best model. |
| | $\hat{b}$ | 3.12 | | | | | |
| Rayleigh | | | 0 | -20295 | <0.000001 (1E-6) | 197781 | |
| K | $\hat{\alpha}$ | 1.59 | 0.129 | -1673 | <0.000001 (1E-6) | 160539 | |
| Gamma | $\hat{\alpha}$ | 2.14 | 0.438 | -1496 | <0.000001 (1E-6) | 160186 | |
| | $\hat{\beta}$ | 2.69 | | | | | |
| **Intensity Distribution** | **Parameters** | **Parameter Estimates** | **KS p-value** | **LRT R** | **LRT p-value** | **AIC** | **Interpretation** |
| Lomax | $\hat{\lambda}$ | 1.13 | 0.913 | | | 33113 | - KS, LRT, and AIC indicate Lomax is the best model. |
| | $\hat{b}$ | 3.12 | | | | | |
| Exponential | | | 0 | -355891 | <0.000001 (1E-6) | 744890 | |
| K | $\hat{\alpha}$ | 1.59 | 0.215 | -27534 | <0.000001 (1E-6) | 88178 | |
| Gamma | $\hat{\alpha}$ | 2.14 | 0.529 | -30650 | <0.000001 (1E-6) | 94412 | |
| | $\hat{\beta}$ | 2.69 | | | | | |

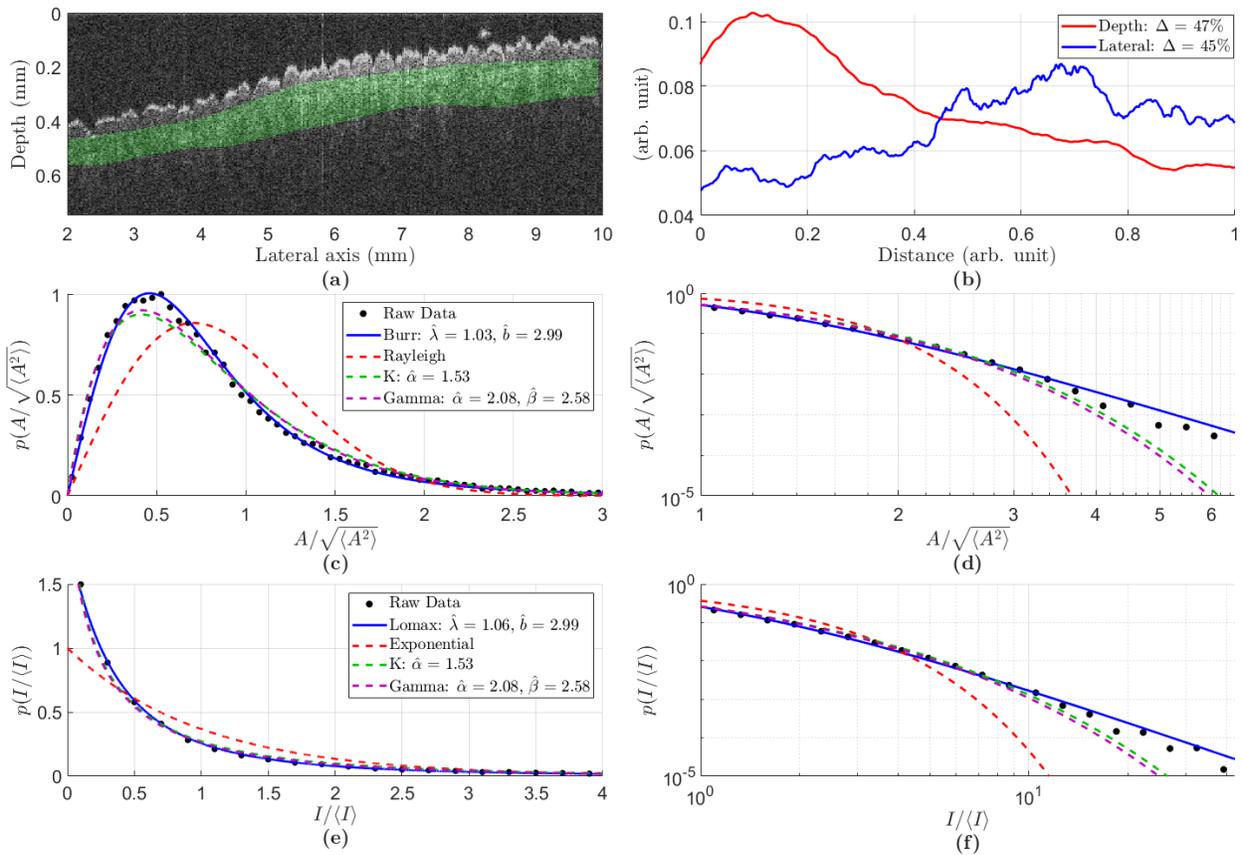

**Figure S9**: Human hand (backside) scanned with 1000 A-lines. **(a)** Single unfiltered and unaveraged B-mode image with ROI shaded green. **(b)** ROI's normalized spatial integration profiles which do not exceed a Δ of 50%. **(c)** MLE fits to normalized amplitude with the Burr, Rayleigh, K, and Gamma distributions on a linear scale with linear binning. **(d)** Same as part (c) except on a log-log scale with logarithmic binning to visualize tail behavior. **(e)** MLE fits to normalized intensity with the Lomax, Exponential, K, and Gamma distributions on a linear scale with linear binning. **(f)** Same as part (e) except on a log-log scale with logarithmic binning to visualize tail behavior.

**Table S9**: Parameters and statistics for the human hand's backside.

| Amplitude Distribution | Parameters | Parameter Estimates | KS p-value | LRT $R$ | LRT p-value | AIC | Interpretation |
|---|---|---|---|---|---|---|---|
| Burr type XII | $\hat{\lambda}$ | 1.03 | 0.947 | | | 58852 | - KS, LRT, and AIC indicate Burr is the best model. |
| | $\hat{b}$ | 2.99 | | | | | |
| Rayleigh | | | 0 | -6090 | <0.000001 (1E-6) | 71028 | |
| K | $\hat{\alpha}$ | 1.53 | 0.495 | -225 | <0.000001 (1E-6) | 59300 | |
| Gamma | $\hat{\alpha}$ | 2.08 | 0.730 | -197 | <0.000001 (1E-6) | 59246 | |
| | $\hat{\beta}$ | 2.58 | | | | | |
| Intensity Distribution | Parameters | Parameter Estimates | KS p-value | LRT $R$ | LRT p-value | AIC | Interpretation |
| Lomax | $\hat{\lambda}$ | 1.06 | 0.863 | | | 15550 | - KS, LRT, and AIC indicate Lomax is the best model. |
| | $\hat{b}$ | 2.99 | | | | | |
| Exponential | | | 0 | -13406 | <0.000001 (1E-6) | 42358 | |
| K | $\hat{\alpha}$ | 1.53 | 0.446 | -1774 | <0.000001 (1E-6) | 19098 | |
| Gamma | $\hat{\alpha}$ | 2.08 | 0.516 | -2087 | <0.000001 (1E-6) | 19726 | |
| | $\hat{\beta}$ | 2.58 | | | | | |

**Table S10:** Summary table of all statistical tests and measures for all tissue samples in order to determine the best distribution. Largest KS *p*-values and smallest AIC value are highlighted in yellow. Rows for LRT *p*-values are not shown if all three values were less than 1E-3.

| Sample | Data | Test or Metric | Burr/ Lomax | Rayleigh/ Exponential | K | Gamma | Best Model |
|---|---|---|---|---|---|---|---|
| 5% Gelatin Phantom (- Milk) | Amplitude | KS *p*-value | 0.964 | 0 | 0.449 | 0.762 | Burr |
| | | LRT sign($R$) | | - | - | - | |
| | | AIC | 38823 | 46432 | 39614 | 39499 | |
| | Intensity | KS *p*-value | 0.909 | 0 | 0.428 | 0.720 | Lomax |
| | | LRT sign($R$) | | - | - | - | |
| | | AIC | 19255 | 64787 | 25540 | 26278 | |
| 5% Gelatin Phantom (+ Milk) | Amplitude | KS *p*-value | 0.958 | 0.033 | 0.924 | 0.841 | Burr |
| | | LRT sign($R$) | | - | - | - | |
| | | AIC | 222575 | 231145 | 222850 | 223528 | |
| | Intensity | KS *p*-value | 0.869 | 0.002 | 0.738 | 0.095 | Lomax |
| | | LRT sign($R$) | | - | - | - | |
| | | AIC | 20250 | 45842 | 24577 | 22355 | |
| Mouse Brain | Amplitude | KS *p*-value | 0.949 | 0.075 | 0.929 | 0.816 | Burr |
| | | LRT sign($R$) | | - | - | - | |
| | | AIC | 91077 | 94317 | 91210 | 91614 | |
| | Intensity | KS *p*-value | 0.851 | 0.036 | 0.828 | 0.662 | Lomax |
| | | LRT sign($R$) | | - | - | - | |
| | | AIC | 16400 | 28956 | 18389 | 16906 | |
| Mouse Liver | Amplitude | KS *p*-value | 0.957 | 0.043 | 0.935 | 0.879 | Burr |
| | | LRT sign($R$) | | - | - | - | |
| | | LRT *p*-value | | < 1E-6 | 0.091 | < 1E-6 | |
| | | AIC | 65863 | 68220 | 65898 | 66131 | |
| | Intensity | KS *p*-value | 0.841 | 0.004 | 0.771 | 0.223 | Lomax |
| | | LRT sign($R$) | | - | - | - | |
| | | AIC | 19147 | 39990 | 22551 | 24554 | |
| Pig Brain | Amplitude | KS *p*-value | 0.892 | 0.013 | 0.943 | 0.913 | K |
| | | LRT sign($R$) | | - | + | + | |
| | | LRT *p*-value | | < 1E-6 | < 1E-6 | 0.15 | |
| | | AIC | 9696 | 9941 | 9635 | 9676 | |
| | Intensity | KS *p*-value | 0.886 | 0.005 | 0.925 | 0.921 | K |
| | | LRT sign($R$) | | - | + | + | |
| | | LRT *p*-value | | < 1E-6 | 0.024 | < 1E-6 | |
| | | AIC | 8824 | 10444 | 8575 | 8834 | |
| Pig Cornea | Amplitude | KS *p*-value | 0.958 | 0 | 0.391 | 0.626 | Burr |
| | | LRT sign($R$) | | - | - | - | |
| | | AIC | 134916 | 165495 | 136840 | 136641 | |
| | Intensity | KS *p*-value | 0.944 | 0 | 0.432 | 0.697 | Lomax |
| | | LRT sign($R$) | | - | - | - | |
| | | AIC | 20650 | 103331 | 31193 | 32619 | |
| Chicken Muscle | Amplitude | KS *p*-value | 0.952 | 0.001 | 0.884 | 0.899 | Burr |
| | | LRT sign($R$) | | - | - | - | |
| | | AIC | 30196 | 32621 | 30340 | 30314 | |
| | Intensity | KS *p*-value | 0.907 | 0 | 0.845 | 0.876 | Lomax |
| | | LRT sign($R$) | | - | - | - | |
| | | AIC | 22988 | 79951 | 31863 | 30425 | |
| Human Hand (Palm) | Amplitude | KS *p*-value | 0.894 | 0 | 0.129 | 0.438 | Burr |
| | | LRT sign($R$) | | - | - | - | |
| | | AIC | 157194 | 197781 | 160539 | 160186 | |
| | Intensity | KS *p*-value | 0.913 | 0 | 0.215 | 0.529 | Lomax |
| | | LRT sign($R$) | | - | - | - | |
| | | AIC | 33113 | 744890 | 88178 | 94412 | |
| Human Hand (Back) | Amplitude | KS *p*-value | 0.947 | 0 | 0.495 | 0.730 | Burr |
| | | LRT sign($R$) | | - | - | - | |
| | | AIC | 58852 | 71028 | 59300 | 59246 | |
| | Intensity | KS *p*-value | 0.863 | 0 | 0.446 | 0.516 | Lomax |
| | | LRT sign($R$) | | - | - | - | |
| | | AIC | 15550 | 42358 | 19098 | 19726 | |